\documentclass[useAMS,usenatbib,usefloat]{mnras}
\usepackage{float}
\usepackage{amsmath}
\usepackage{amssymb}
\usepackage{rotating}
\usepackage{tabularx}
\usepackage{gensymb}
\usepackage{longtable}
\usepackage{lscape}

\include{journaldefs}
\usepackage[font=footnotesize,labelfont=bf]{caption}

\begin{document}

\title[\bf The O star hinterland of NGC 3603]{\bf The O star hinterland of the Galactic starburst, NGC 3603}
\author[J. E. Drew et al]{
{\parbox{\textwidth}{J. E. Drew$^{1}$\thanks{E-mail: j.drew@herts.ac.uk}, M. Mongui\'o$^1$, N. J. Wright$^2$}
}\\ \\
$^{1}$School of Physics, Astronomy \& Mathematics, University of Hertfordshire, Hatfield AL10 9AB, UK \\
$^2$Astrophysics Group, Keele University, Keele, ST5 5BG, UK\\
}

\maketitle

\begin{abstract}
The very bright and compact massive young cluster, NGC 3603, has been cited as an example of a starburst in the Milky Way and compared with the much-studied R136/30 Doradus region in the Large Magellanic Cloud. Here we build on the discovery by \cite{MMS2017} of a large number of reddened O stars around this cluster.  We construct a list of 288 candidate O stars with proper motions, in a region of sky spanning $1.5\times1.5$ square degrees centered on NGC 3603, by cross-matching the Mohr-Smith et al. (2017) catalogue with Gaia DR2 \citep{Gaiadr2}.  This provides the basis for a first comprehensive examination of the proper motions of these massive stars in the halo of NGC 3603, relative to the much better studied central region.  We identify up to 11 likely O star ejections -- 8 of which would have been ejected between 0.60 and 0.95 Myr ago (supporting the age of $\sim$1~Myr that has been attributed to the bright cluster centre). Seven candidate ejections are arranged in a partial ring to the south of the cluster core spanning radii of 9--18 arcmin (18--36 pc if the cluster is 7 kpc away).   We also show that the cluster has a halo of a further $\sim$100 O stars extending to a radius of at least 5 arcmin, adding to the picture of NGC 3603 as a scaled down version of the R136/30 Dor region.
\end{abstract}



\section{Introduction}

Massive young clusters are rare objects that nevertheless exert a major influence on their galactic environments.  Their modes of formation remain an important topic of research, echoing continuing significant uncertainty in the modes of evolution of the most massive stars that are their distinctive constituents.  Clues to the early internal dynamics of massive young clusters can come from the O stars they eject \citep{Fujii2011}.  With the arrival of Gaia DR2 proper motions and the availability of wide field photometric surveys, the opportunity now exists to locate ejected O stars in the environs of their birth clusters.  In a previous paper \citep{Drew2018} we studied the example of Westerlund 2 and discovered a surprisingly ordered 'twin-exhaust' pattern of ejections that may favour the creation of Westerlund 2 via the merger of distinct sub-clusters.  Here we move on to the example of NGC 3603, the even {\bf brighter and more compact} clustering in the region.

Like Westerlund 2, NGC 3603 is in the Carina region of the Galactic Plane.  It is one of a small number of very dense and massive clusters in the Milky Way.  Indeed, the extreme stellar density in the core has prompted comparisons with the extragalactic starburst phenomenon \citep[e.g.][]{Eisenhauer1998,Moffat2002,Stolte2006}.  The mass of the cluster is among the highest measured in the Milky Way:  \cite{Harayama2008} placed it in the range from 10\,000 up to 16\,000 M$_{\odot}$, while \cite{Rochau2010} obtained $\sim$18\,000 M$_{\odot}$.  The age often cited for NGC 3603, based on the stellar content of its inner core of diameter $\sim20$~arcsec is 1 to 2 Myr \citep[e.g.][]{Sung2004,Melena2008,Kudryavtseva2012}.  This core is sometimes referred to as the HD 97950 cluster.    Measurements over a wider sky area out to a radius of an arcminute or so, have indicated that an older, lower density population may be present as well \citep{Melena2008,Beccari2010}.

In earlier work \citep[][hereafter MS-I and MS-II]{MMS2015, MMS2017}, we presented blue selections of OB stars from the VST 
Photometric H$\alpha$ Survey of the Southern Galactic Plane and Bulge \citep[VPHAS+,][]{Drew2014} across the Carina region, and showed that their conversion to spectroscopically-confirmed OB stars is very high.  The MS-II list of 5915 O-B2 candidates is accompanied by high-quality measures of extinction, along with estimates of effective temperatures that are good enough to broadly classify as early-O, later-O and early-B stars. These are derived from fitting each object's spectral energy distribution (SED) as represented by its $u/g/r/i/J/H/K$ magnitudes.  Here, we reuse MS-II in order to focus on the hinterland of NGC 3603.  

This paper is organised as follows.  First, we select from MS-II a set of high-confidence O stars within a region of 1.5$\times$1.5 sq.deg centred on NGC 3603, and crossmatch it with the Gaia DR2 database \citep{Gaiadr2} with the aim of utilising the proper motion (PM) data (Section~\ref{sec:sample}).  We then compute the mean proper motion of the cluster using stars located within 1 arcminute of the cluster centre which becomes the basis for relative proper motions (rPM, see sections~\ref{sec:core-pm} and \ref{sec:rel-pms}).  This enables the identification of probable cluster escapes that turn out, intriguingly, to be located in a half ring (section \ref{sec:ejections}).  After some sample decontamination, we consider the radial distribution of the O star candidates and find it largely follows the King model adopted by \cite{Harayama2008}.  The paper ends with a discussion of the significance of the results and some conclusions (sections~\ref{sec:discussion} and \ref{sec:conclusions}).  

\section{Construction of the sample}

\label{sec:sample}

\begin{figure}
\begin{center}
\includegraphics[width=0.85\columnwidth]{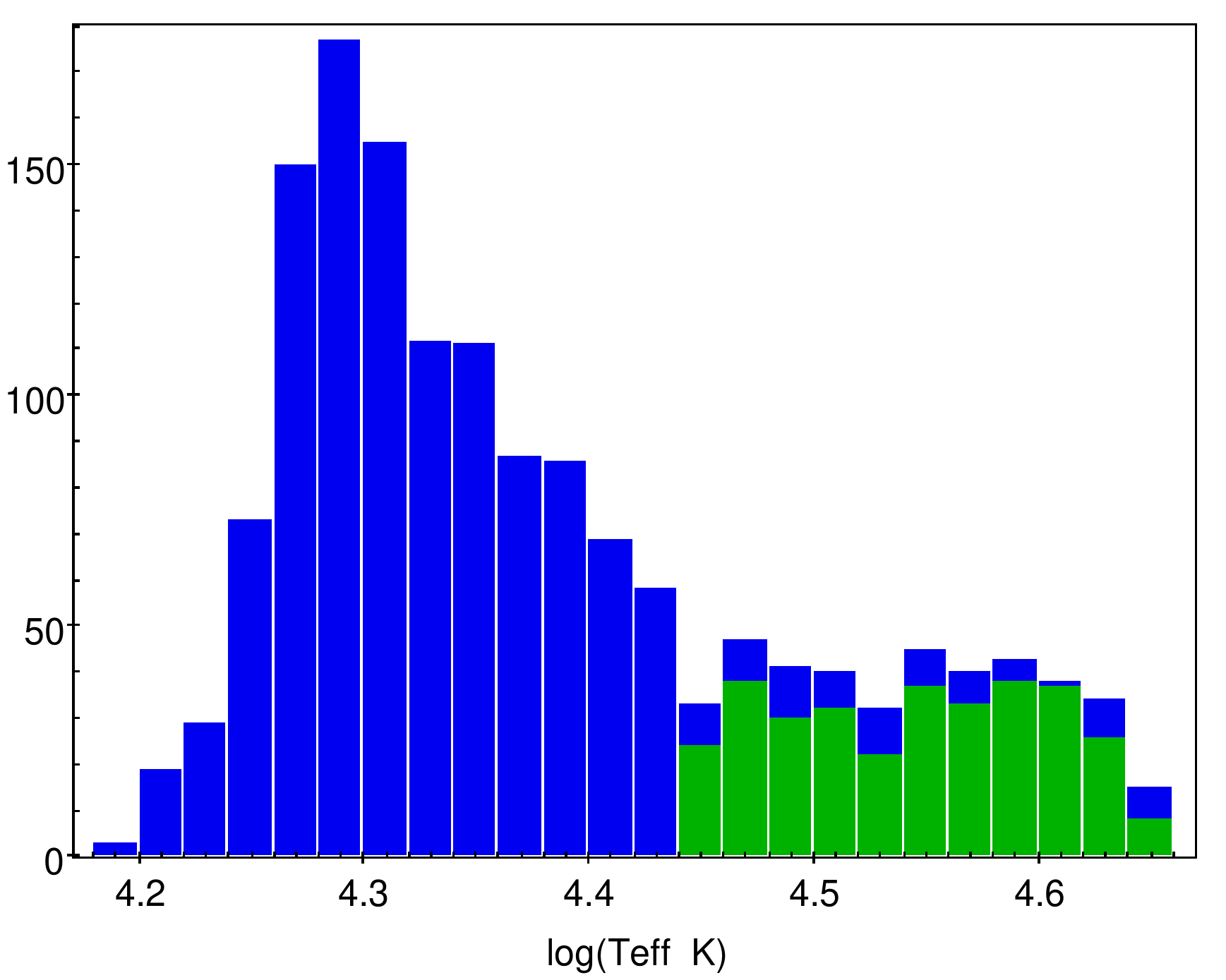}
\includegraphics[width=0.85\columnwidth]{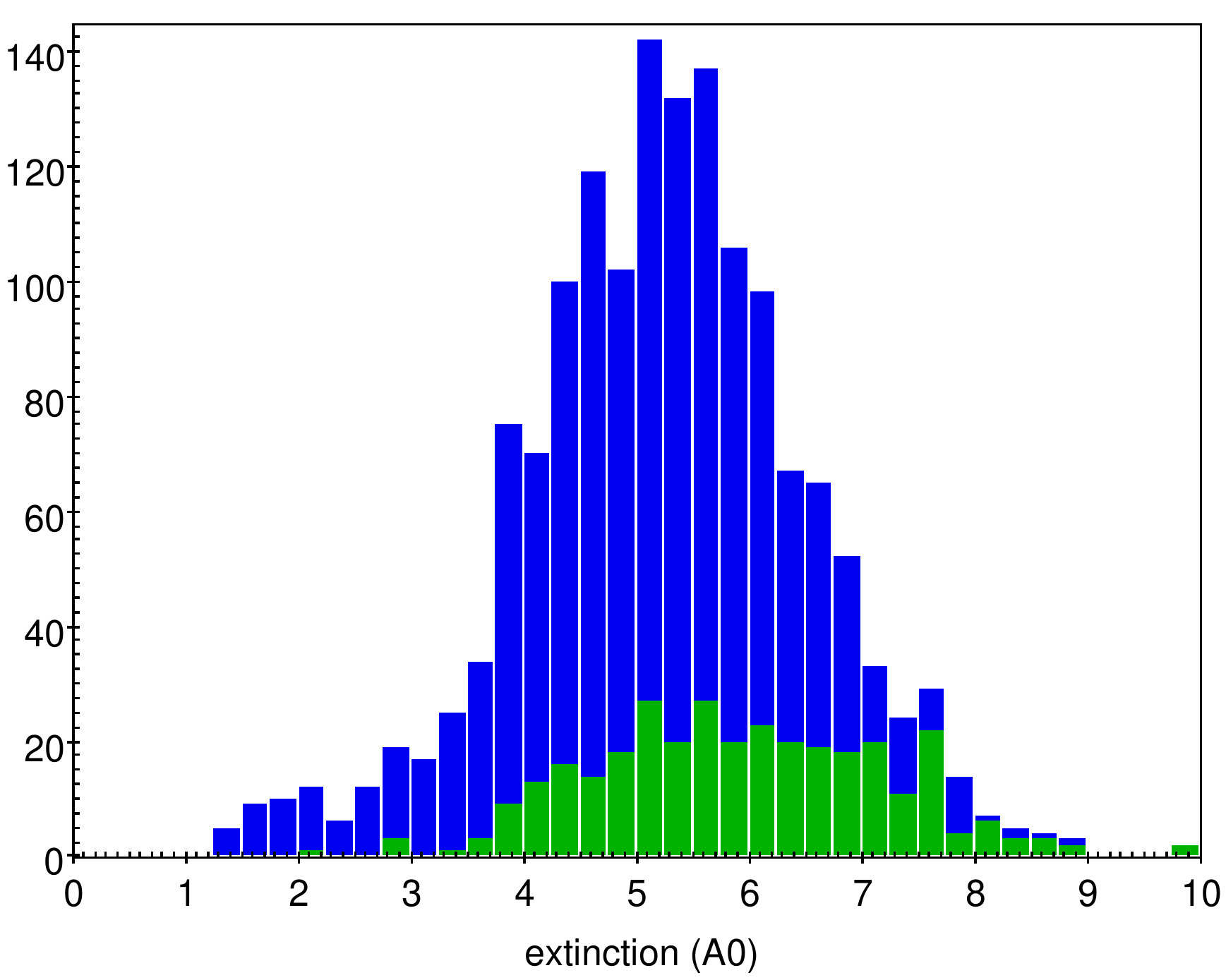}
\includegraphics[width=0.85\columnwidth]{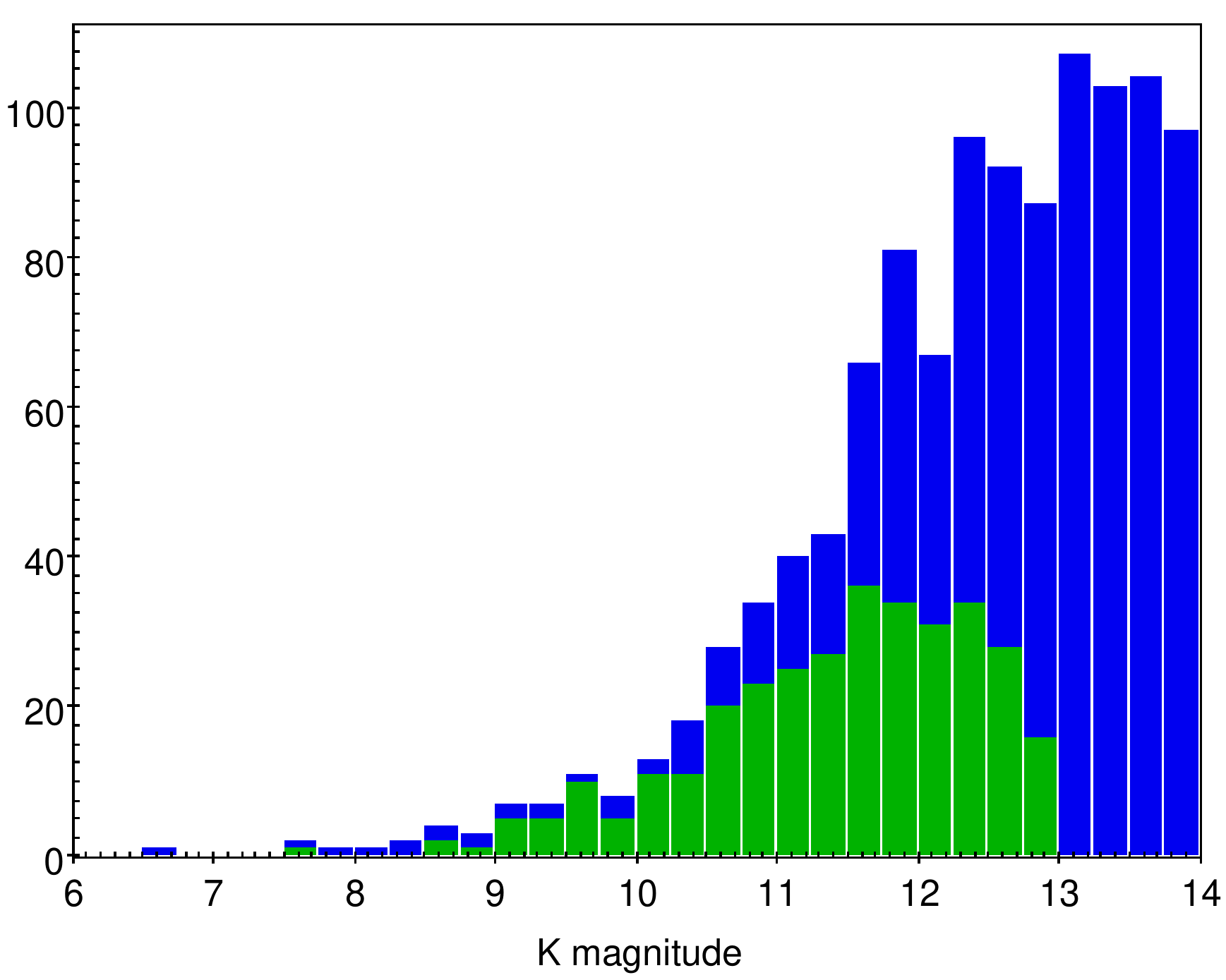}
\caption{
These histograms show how the MS-II objects inside the 1.5$\times$1.5 deg$^2$ box centred on NGC 3603, satisfying $\chi^2<25$, are distributed in effective temperature (top panel), extinction, $A_0$ (middle) and $K$ magnitude (bottom).  The full sample of 1537 stars is shaded in blue.  The green bars pick out the selection with logT$_{\rm eff}$ $> 4.44$ and $K < 13.0$ mag.  These are the focus of this study.
}
\label{fig:properties}
\end{center}
\end{figure}

We adopt as the central reference position in the core of NGC 3603 the Galactic coordinates, $\ell = 291^{\circ}.617$, $b = -0^{\circ}.523$.  The selection box on the sky around this position is a square occupying 1.5$\times$1.5~deg$^2$, with sides oriented along the directions of constant Galactic longitude and latitude.  Within this region, the MS-II database provides a list of 1663 blue-selected OB candidate stars.  Of these we retain as a long list those objects for which the reported quality of fits to their optical-NIR SEDs is $\chi^2 < 25$.  MS-II recommended a tighter $\chi^2$ limit than this for the selection of "good OB stars": it is relaxed here so as not to exclude detected objects close to the centre of NGC 3603 that are known already to be early-type emission line stars.   This reduces the list to 1537 objects.  How these stars are distributed in terms of best-fit effective temperature (or $\log(T_{\rm eff} K)$ as plotted) and extinction, $A_0$, and 2MASS $K$ magnitude is shown in Figure~\ref{fig:properties}.  We remind that $A_0$ represents the monochromatic extinction at a wavelength of 5495~\AA .

To focus the long list down onto the likely O star content of the region at a distance $D \lesssim 8$ kpc, and extinction $A_0 \lesssim 10$ magnitudes, we cut both on effective temperature such that $\log(T_{\rm eff} K) > 4.44$ and on $K$ magnitude such that $K < 13.0$.  The first cut admits objects with estimated effective temperatures greater than 27500~K: since O stars are associated with effective temperatures exceeding $\sim$30000~K, this builds in some margin for fit error.  The reasoning behind the second cut is that, since $M_K \simeq -3$ for an O9.5 main sequence star \citep{Martins2005}, such a star at a maximal distance of 8 kpc, with $A_0 = 10$, would suffer no more than $\sim 1$ magnitude of $K$ extinction, resulting in an apparent magnitude of $K \sim 12.5$. Setting the fainter limit of $K = 13.0$ again makes some allowance for error.  How the reduced sample compares with the full list from MS-II is illustrated in Figure~\ref{fig:properties}.

\begin{figure*}
\begin{center}
\includegraphics[width=1.4\columnwidth]{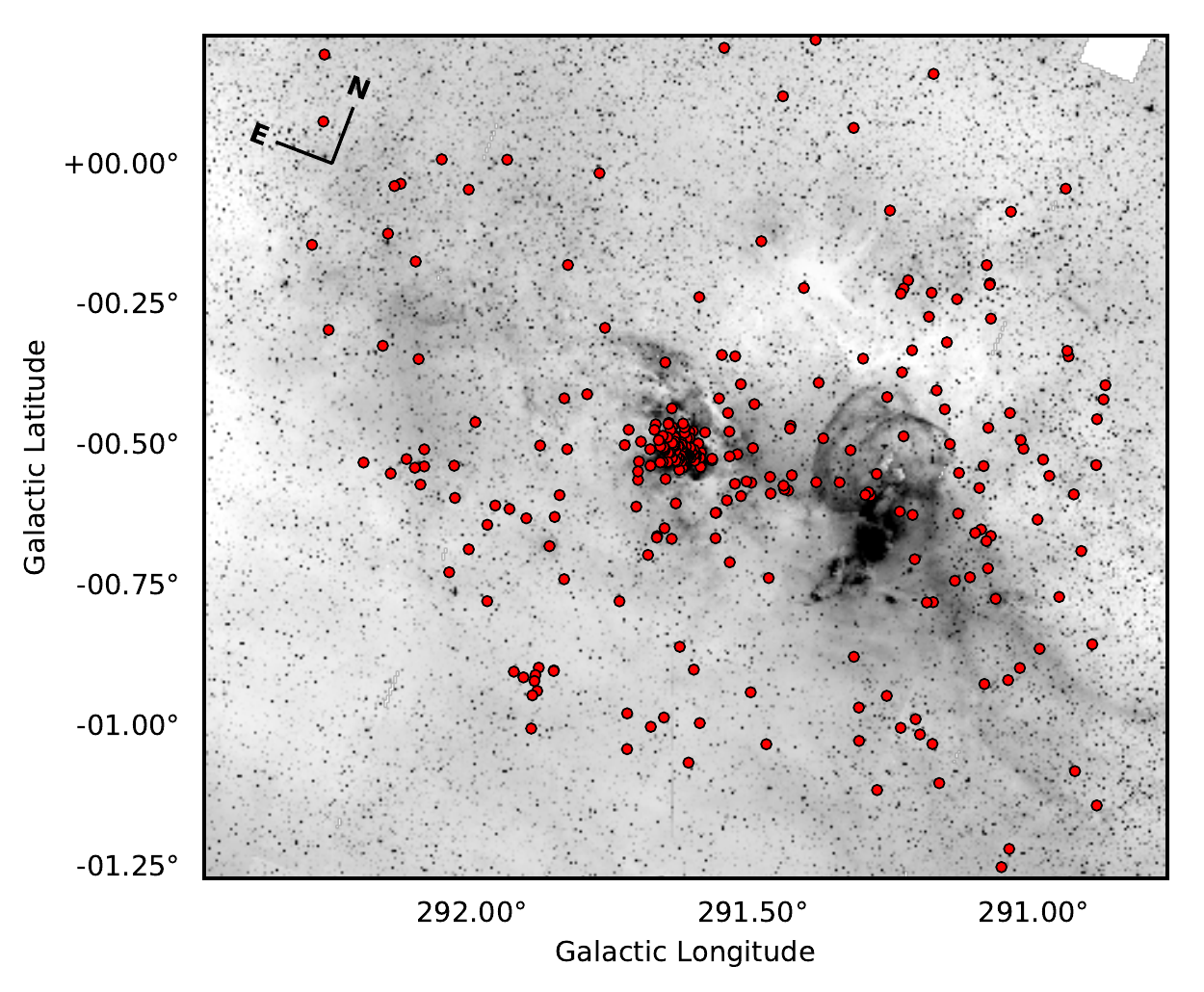}
\caption{
The on-sky positions of the 288 selected objects in the region around NGC 3603.  The background image is constructed from VPHAS+ $H\alpha$ data.  
}
\label{fig:crossmatch}
\end{center}
\end{figure*}

Next, a cross-match of the reduced list of 325 stars with the Gaia DR2 catalogue was undertaken.  We found that 292 objects with good astrometry were successfully matched up at position offsets ranging from $\sim$0.03 to $\sim$0.2 arcsec (with a mean of $\sim0.1$ arcsec).   The astrometry is "good" in that all of them pass the mission-recommended test of astrometric quality: that the renormalised unit weight error, $u/u_0(G,BP-RP) < 1.4$, where $u = \sqrt{(\chi^2)/(N-5)}$ and $u_0(G,BP-RP)$ is the magnitude- and colour-dependent reference value.\footnote{See the Gaia mission document, GAIA-C3-TN-LU-LL-124-01 by L. Lindegren}  The minimum number of Gaia satellite visibility periods per source is 13, while the median value is 18.  As might be expected, given the centering of the Gaia $G$ transmission in the red part of the optical spectrum, the typical difference between MS-II $r$ and Gaia $G$ magnitudes per source is modest, being in the region of 0.1-0.2 mag.  The apparent magnitudes of the selection fall mainly in the range, $12.0 < r, G < 17.5$. 

Finally, we removed 4 objects that are likely to be well into the foreground of NGC 3603 -- specifically, stars with $>5\sigma$ Gaia DR2 parallaxes for which distances of under 4 kpc are indicated.  This leaves 288 candidate O stars of which 15 are already in the literature.  In the online appendix, a table of names, positions and other important quantities is provided.  This is our sample. How it is distributed on the sky is shown in Figure~\ref{fig:crossmatch}.   

The absence of candidate O stars towards higher Galactic longitudes and more negative latitudes apparent in figure~\ref{fig:crossmatch} is real, in the sense that the MS-II source catalogue only found cooler B stars in this corner (along with 4 indeterminate objects returning unacceptably high $\chi^2$ SED fits).  Another prominent feature of the emerging distribution is the dense, extended core of objects around the cluster position.  It is a reasonable first conjecture that most of the objects are associated with NGC 3603: within 2 arcminutes of our fiducial position there are 56 O-star candidates, rising to 115 inside a radius of 8 arcminutes.

Within the field under consideration there could be a concern that there will be 'contamination' due to objects located in the first crossing of the Carina Arm at a distance of 2--3 kpc (with NGC 3603 in the second crossing at $>6$ kpc, see section~\ref{sec:distance}).  A sign of this in Figure~\ref{fig:crossmatch} is the presence of the bright nebula, NGC 3576, around $\ell \simeq 291^{\circ}.2$, $b \simeq -0^{\circ}.65$.  This is known to be associated with the near Carina Arm \citep[][give a distance of 3.0$\pm$0.3 kpc]{Depree1999}.  Accordingly we could anticipate some of the scatter of objects in the vicinity of this nebula to be near-arm stars.  

However the visual  extinctions of the exciting stars of NGC 3576 itself are high at over 10 magnitudes \citep{Figueredo2002} while the known OB stars in its vicinity are too bright to appear in our sample (e.g. EM Car, an O8V+O8V binary, $V=9.54$).  Furthermore, only 15 of the 288 stars has $A_0 < 4$ -- for which a distance of under $\sim$4 kpc would be implied if $A_0$ were to rise with distance at a rate of 1 mag kpc$^{-1}$ (see the discussion in section 5.3 of MS-II).  We conclude that the amount of near-arm contamination -- which would most likely be dominated by B stars --  is well under 10 percent.


\section{The distance to NGC 3603}
\label{sec:distance}

It is widely accepted that NGC 3603 is located in the far, rather than the near, Carina Arm.     \cite{Melnick1989} presented and analysed UBV photometry of 74 stars over a region of $\sim$9 square arcminutes, determining a distance modulus of 14.3 (or $D = 7.24$ kpc).  \cite{Sung2004} focused mainly on the inner region within 1 arcminute of the cluster centre and used dereddened VI photometry of stars to deduce a distance modulus of $14.2 \pm 0.2$ (or $D = 6.9 \pm 0.6$ kpc).  Since then \cite{Melena2008} have argued for $D = 7.6$ kpc, based on a spectrophotometric analysis of mainly cluster O stars.  Usefully, these authors also provided two tables summarising stellar and kinematic distances in the literature prior to their work (see their Tables 4 and 5).  A kinematic measure not included in this compilation was that by \cite{Nurnberger2002} based on CS line observations: they obtained $D=7.7\pm0.2$ kpc, from CS velocities of $14.2\pm 1.6$ km s$^{-1}$ (LSR).  
The emergent picture from the literature is that since the mid 1980s most estimates have ranged from 6 kpc up to 8 kpc.  Here, we will work with $D = 7\pm 1$ kpc.

\begin{figure}
\begin{center}
\includegraphics[width=0.8\columnwidth]{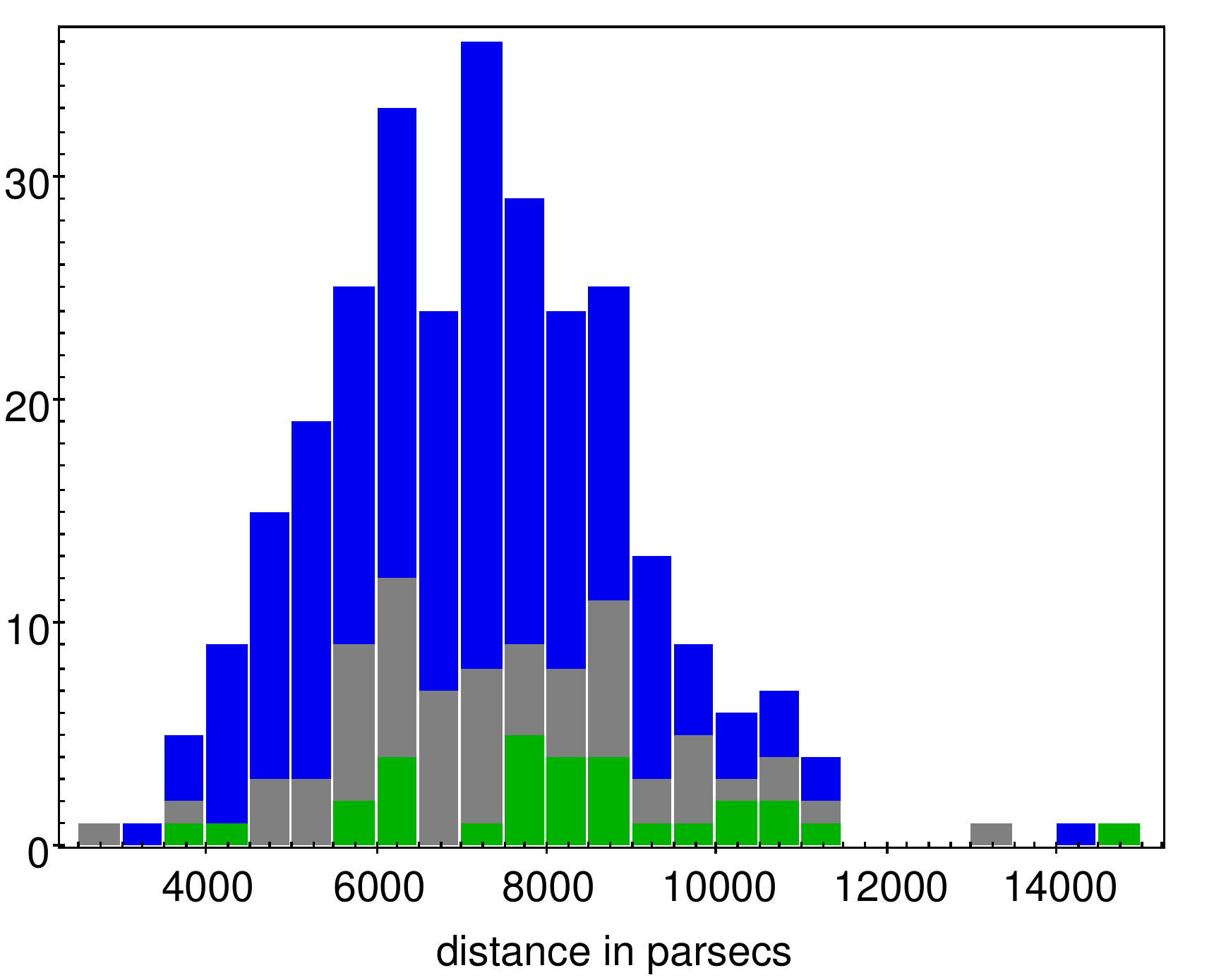}
\caption{
The histogram of distances inferred from Gaia DR2 parallaxes for the 288 selected O-star candidates, using an EDSD prior with scale length of 1.5 kpc.  Every parallax was increased by 0.03 mas to allow for the known Gaia DR2 global offset.  The blue histogram bars represent the entire sample.  The grey bars are the distribution obtained on limiting the selection to stars within 5 arcmin of the cluster fiducial position, while the green colour picks out stars within 1 arcmin.
}
\label{fig:edsd-distance}
\end{center}
\end{figure}

A distance of 7 kpc to NGC 3603 implies an astrometric parallax of 0.143 mas would be measured in the ideal case of negligible error.  The Gaia DR2 results are known to exhibit an offset of -0.03 mas, and a position dependent systematic error of up to 0.1 mas \citep{Lindegren2018}.  In this situation, we should not expect Gaia DR2 parallax data to do more than perhaps statistically corroborate existing measures of distance. Figure~\ref{fig:edsd-distance} shows the distribution of distances for our sample of 288 stars inferred on applying the EDSD (exponentially-decreasing space density) prior described by \cite{Luri2018}: a scale length of $L = 1.5$~kpc was adopted and 0.03 mas was added to each parallax to correct for the global Gaia DR2 offset.  

Evidently the distribution in Figure~\ref{fig:edsd-distance} is very broad.  This remains true when the selection is limited to those close to cluster centre.  This is unsurprising in itself.  What is of more interest is the trend in the mean and median measures as the sample is reduced from the full set to, first, projected separations from the centre of $<5$ arcmin, and then to $<1$ arcmin.  The changes are in the sense of increased mean, or median, distance as the sky area is restricted: the means and formal errors from the distributions are
\begin{itemize}
\item full sample: $D=7.2\pm0.1$ kpc (288 stars),
\item within 5 arcmin of centre: $D=7.6\pm0.2$ kpc (93 stars),
\item within 1 arcmin of centre; $D=8.2\pm0.4$ kpc (30 stars).
\end{itemize}
The medians for these three samples are 7.2, 7.5 and 8.1 kpc respectively -- i.e. scarcely different from, if a little lower than the means.

This pattern is plausible even if the numerical values of the distances are subject to a presently undetermined systematic error (capable of shifting the means by more than 1 kpc either way).  We would expect the MS-II catalogue to provide more candidates in the foreground of NGC 3603 than in the increasingly reddened background, resulting in a bias toward a shorter distance in the largest sample.  This is present, but it is not strong as it only indicates a difference in the mean of around 1 kpc ($\sim 15$ percent) between the stars in the wider environment and those in the cluster centre.  

Qualitatively, the outcome of this exercise is not sensitive to the EDSD scale length prior: for example, if it is doubled to 3 kpc there is again a gradual rise in the mean distance estimate as the sky area sampled is focused more onto NGC 3603.

\section{Results}

\subsection{The proper motion of the core of NGC 3603}
\label{sec:core-pm}

\begin{figure}
\includegraphics[width=1.15\columnwidth]{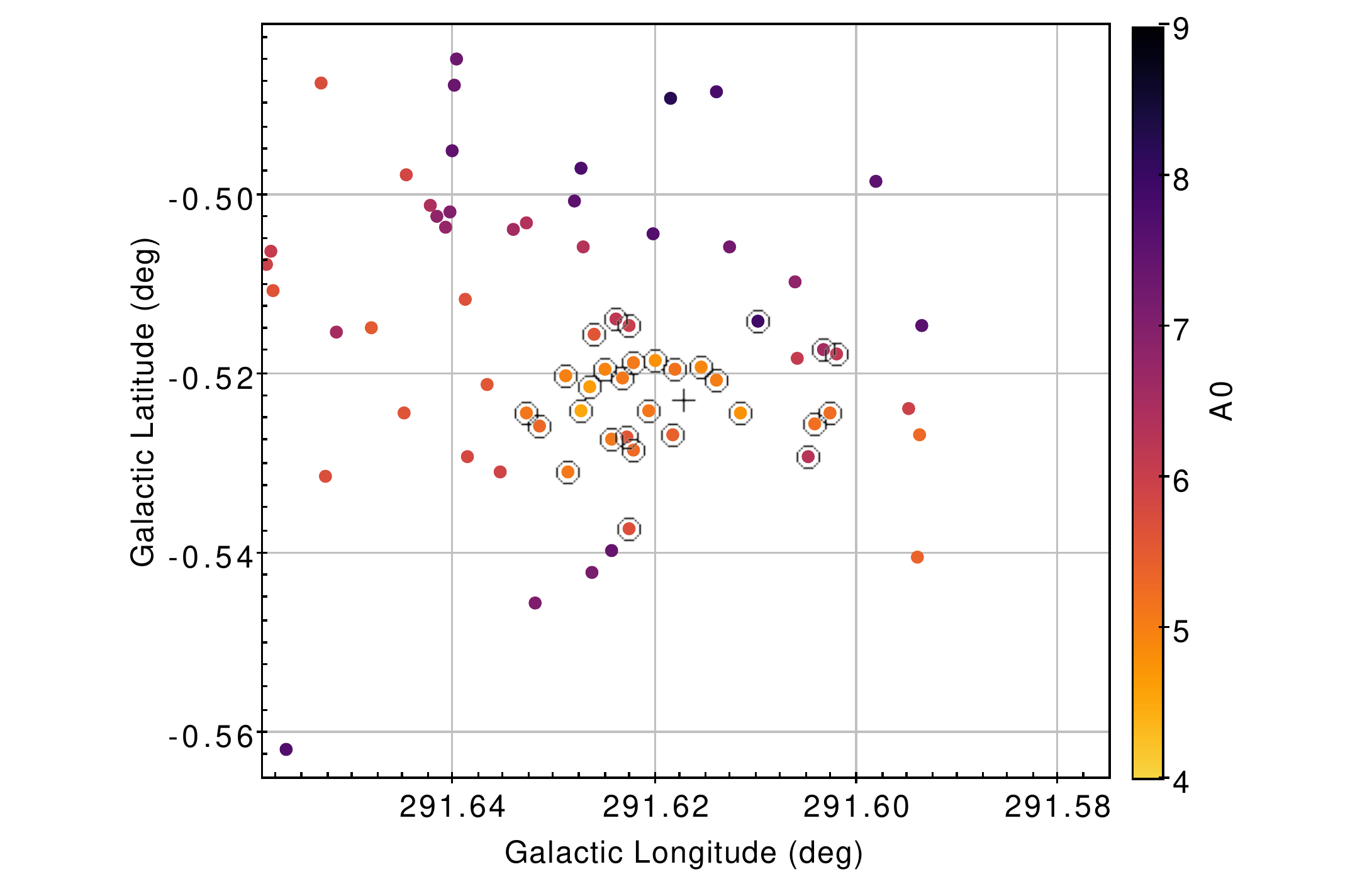}
\centering
\caption{A zoom into a $5\times5$arcmin$^2$ box around the centre of the cluster (it contains 69 stars).  Because of the extreme confusion in the stellar core as imaged from the ground, no object lies within 12 arcsec of the centre, which is marked with a plus symbol.  The 29 stars within a radius of 1 arcmin that were used to determine the mean cluster proper motion are encircled. All objects are coloured according to extinction.
}
\label{fig:zoom}
\end{figure}

\begin{figure}
\begin{center}
\includegraphics[width=0.95\columnwidth]{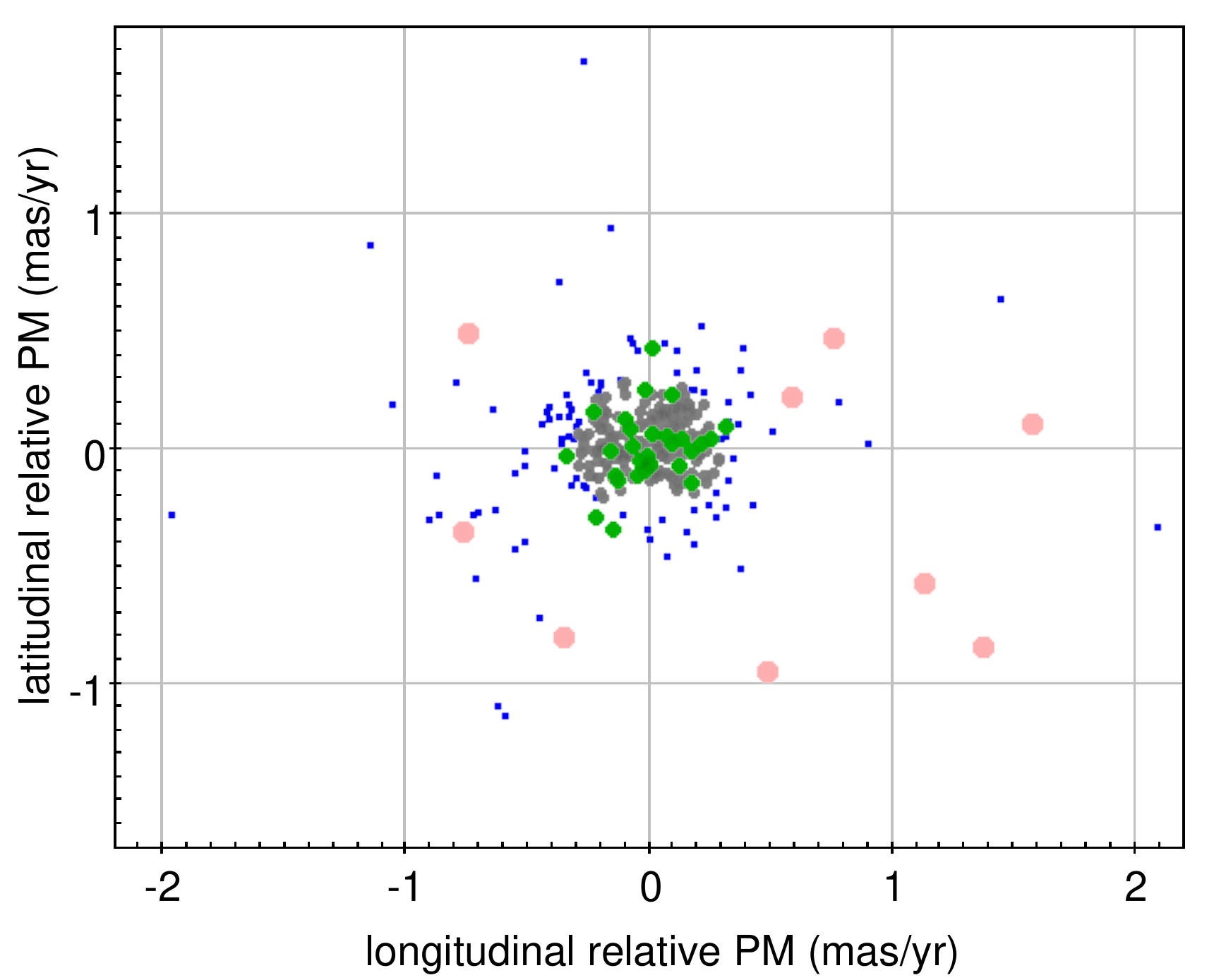}
\includegraphics[width=0.95\columnwidth]{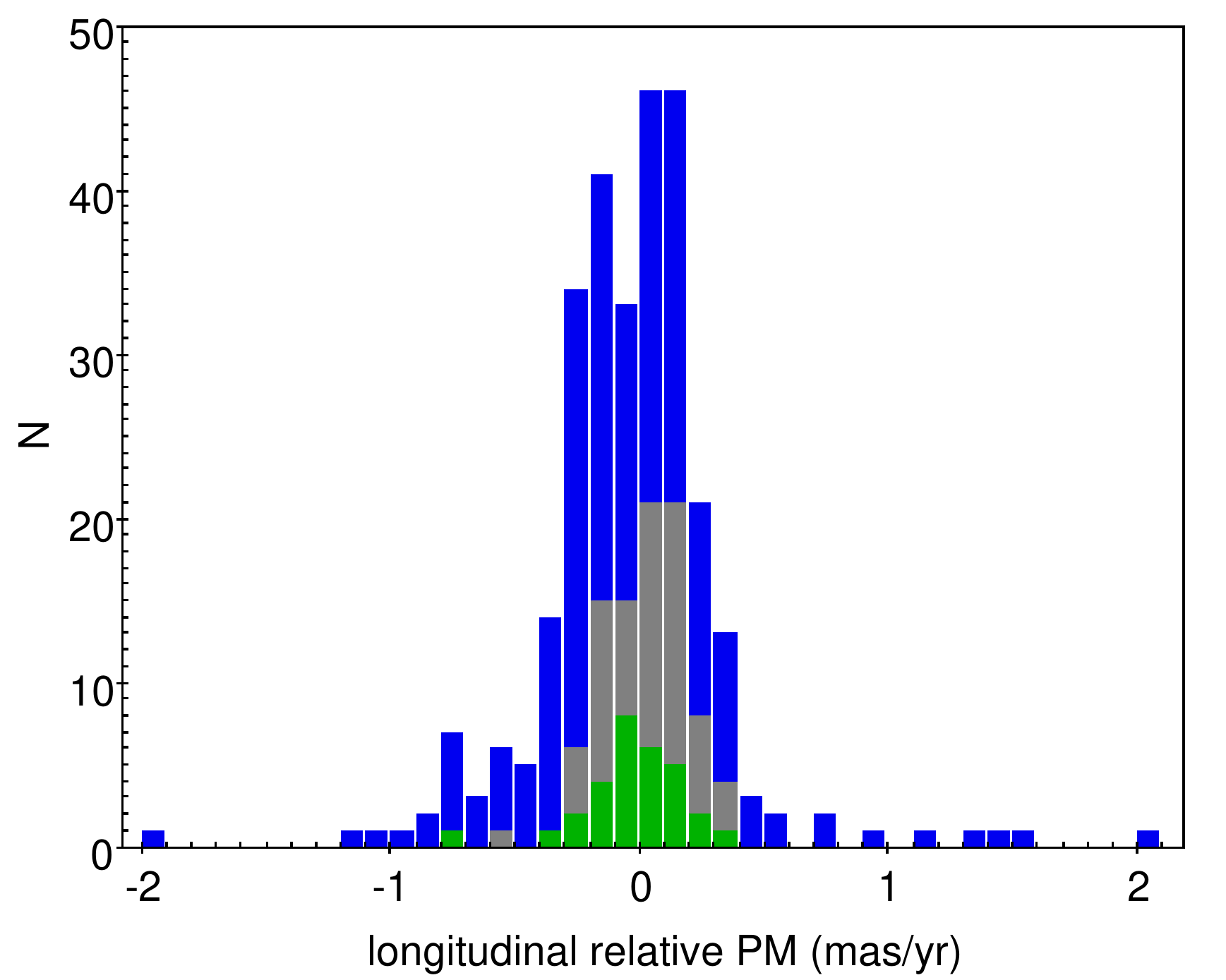}
\caption{The proper motions of the retained sample of objects relative to the core of NGC 3603 in mas yr$^{-1}$.  In the upper panel, the 29 stars within 1 arcmin of the cluster centre used to compute the cluster PM are shown in green.  Stars in grey have relative proper motions of magnitude less than 0.3 mas yr$^{-1}$ -- they can be seen as commensurate with those typifying the core region.  62\% of the sample (180 stars) fall in this group. The 9 objects in pink are candidate ejections with relative PMs exceeding 0.6 mas yr$^{-1}$ and trajectories passing within 1 arcmin of the centre of NGC 3603.  All other stars are in blue.  The lower panel is the histogram of the longitude component of the relative PM for all objects (in blue).  Notice the double-peaked character of the longitude distribution and the negative shoulder showing between $-0.4$ and $-0.8$ mas yr$^{-1}$.  For comparison the distributions obtained on limiting the selection to within 5 arcmin (grey) and 1 arcmin of cluster centre (green) are superposed: the shoulder and double-peaking weaken and disappear. 
}
\label{fig:relpms}
\end{center}
\end{figure} 

There are 30 cross-matched stars with astrometry, accepted into the sample, that lie within 1 arcminute of our fiducial position (see Figure~\ref{fig:zoom}).  But none are closer to the nominal centre than 0.2 arcmin, as a consequence of the severe source confusion in the brilliant cluster core typically seen in ground-based images.  After one evident outlying object with a high relative proper motion is removed from the subset, the remaining 29 stars yield a mean PM in Galactic co-ordinates of $\mu_{\ell,*} = -5.881 \pm 0.151$ mas yr$^{-1}$, and $\mu_b = -0.209 \pm 0.151$ mas yr$^{-1}$.  A check against the 6 Gaia DR2 objects, with no excess astrometric noise located closer to cluster centre, are consistent with this measure.  

At a distance of $\sim$ 7~kpc, the implied mean transverse motion in the $b$ coordinate is equivalent to $\sim$7 km s$^{-1}$, directed below the plane.  This fits with what we would expect if this massive young cluster has negligible vertical motion: the observed $\mu_b$ should then be equal to and opposite in sense to the Sun's motion -- which has indeed been found to be $+$7 km s$^{-1}$ to within one significant figure \citep[see e.g.][]{Schoenrich2010}.

It is also encouraging that the dispersion around the mean we obtain in both coordinates, of 
0.151$\pm$0.020\footnote{The uncertainty is obtained by dividing the standard deviation by $\sqrt{2N-2}$ where $N=29$.} mas yr$^{-1}$, appears consistent with the {\sl Hubble Space Telescope} (HST) results of \cite{Rochau2010} and \cite{Pang2013} based on somewhat fainter intermediate mass stars.  \cite{Rochau2010} obtained an intrinsic 1-D dispersion from 234 stars inside a radius of $\sim0.25$ arcmin of 0.141$\pm$0.027 mas yr$^{-1}$. \cite{Pang2013} considered the same region and a similar scale of sample and measured 0.146$\pm$0.016 and 0.198$\pm$0.016 mas yr$^{-1}$ in two orthogonal directions.

\subsection{Proper motions of the wider sample}
\label{sec:rel-pms}


Proper motions relative to the mean cluster value, $PM_r$, have been computed and are shown in the upper panel of Figure~\ref{fig:relpms}.  There is evidently a tight clustering around the origin that is similar in dispersion to the 'core' objects. Setting an upper bound on the magnitude of the $PM_r$ of 0.3 mas yr$^{-1}$, there would be around 93 qualifying objects that lie within 10 arcmin of cluster centre -- with another 87 with similarly low $PM_r$ scattered across the wider field. 
An important consideration here is whether a small relative proper motion implies proximity and/or dynamical association with NGC 3603.  For some, especially those at modest angular displacements from the cluster, this is likely and credible.  

But we should not forget the other option that some candidate objects are foreground or background and merely tracing Galactic rotation: a distance change of $\sim$1 kpc induces a longitudinal proper motion change of just $\sim$0.2 mas yr$^{-1}$ (depending somewhat on choice of rotation law).  There is evidence this could be happening, as shown in the lower panel of Figure~\ref{fig:relpms}: the distribution in the relative PM longitude component verges on double-peaked and shows a negative shoulder that may signal the mixing in of a higher-PM foreground population.  
When the sample is limited to those within a few arcminutes of the core, the skew and tendency towards double peaking reduces and the shoulder disappears.

\begin{figure}
\begin{center}
\includegraphics[width=0.95\columnwidth]{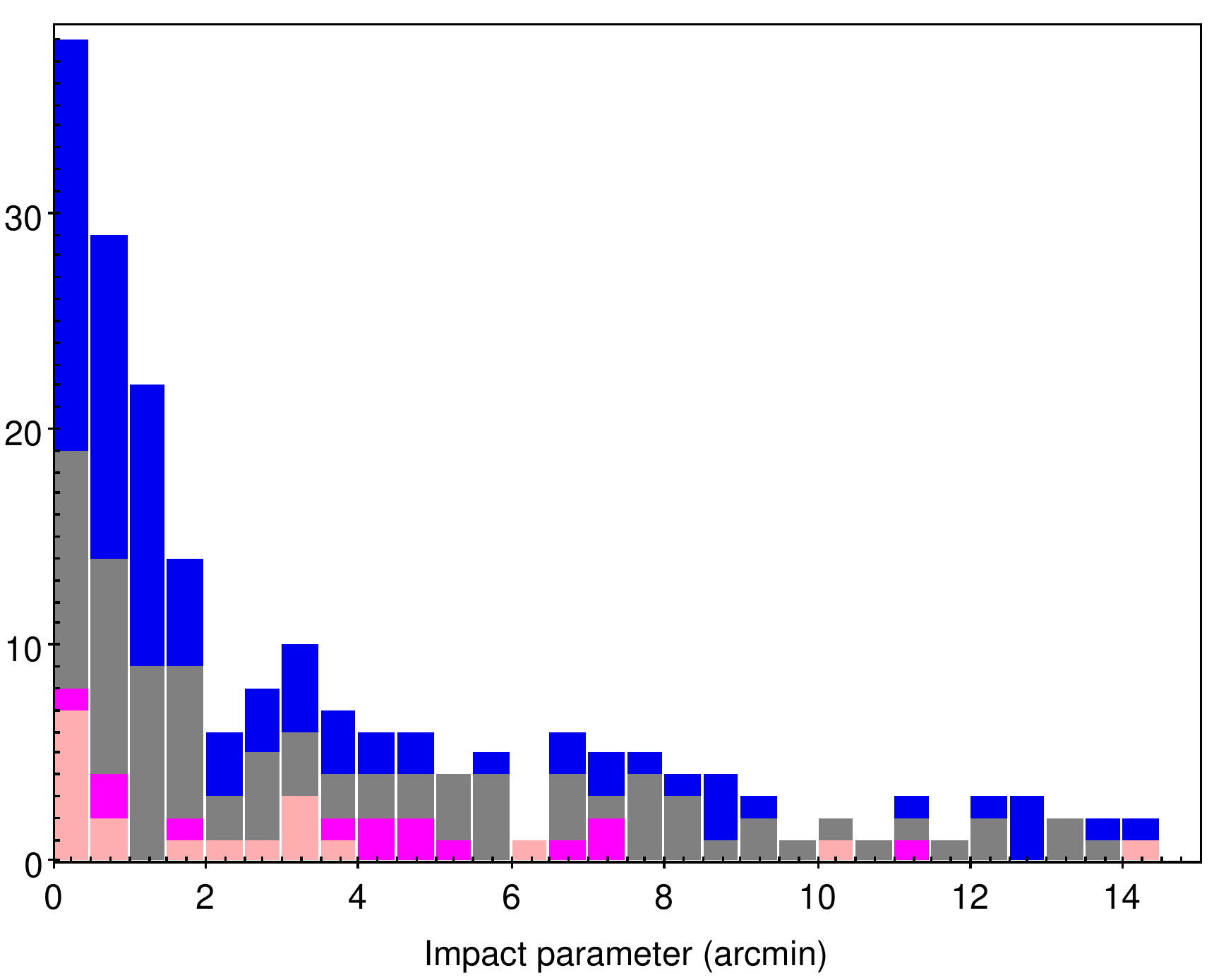}
\caption{
The distribution of computed impact parameters (projected nearest approach to cluster centre) up to 15 arcmin.  The bars are overlaid to pick out different ranges in magnitude of proper motion as follows: blue admits all values; grey signifies objects with $|PM_r| > 0.2$ mas yr$^{-1}$; magenta, $|PM_r| > 0.4$ mas yr$^{-1}$, pink, $|PM_r| > 0.6$ mas yr$^{-1}$. 
}
\label{fig:imppar}
\end{center}
\end{figure}

\subsection{Candidate ejections}
\label{sec:ejections}

\begin{table*}
\caption{Proper motions and related quantities for stars meeting the first two criteria for ejection from the centre of NGC 3603.  Column 1 specifies the MS-II catalogue number. The Gaia DR2 proper motions appear in columns 2 and 3.  Column 4 is the angular distance in arcminutes from the fiducial position $\ell = 291^{\circ}.617$, $b = -0^{\circ}.523$.  Columns 5 and 6 give the proper motion relative to the core-region mean, $PM_r$, in Galactic coordinates.  Column 7 is magnitude of the relative proper motion, while column 8 gives the trajectory impact parameter (in arcminutes).  The (distance-independent) travel time from the fiducial position is in column 9.}
{\centering
\begin{tabular}{lccrrrcccl}
\hline  
MS-II & \multicolumn{2}{c}{Proper motion} & Radius & \multicolumn{2}{c}{Relative PM, $PM_r$} & $|PM_r|$ & $IP$ & Travel time & Note \\
\# & \multicolumn{2}{c}{$\mu_{\alpha,*}$, $\mu_{\delta}$ mas/yr} & arcmin & \multicolumn{2}{c}{$\Delta\mu_{\ell,*}, \Delta\mu_{b}$ mas/yr} & mas/yr & arcmin & Myr & \\
\hline
  13362 & $-$6.400$\pm$0.031 & 1.880$\pm$0.030 &   9.68 &  $-$0.766$\pm$0.043 & $-$0.362$\pm$0.040 &  0.847$\pm$0.021 &  0.27$\pm$0.62 &  0.69 & \\  
  13436 & $-$6.167$\pm$0.053 & 1.315$\pm$0.044 &  12.09 &  $-$0.343$\pm$0.061 & $-$0.803$\pm$0.052 &  0.874$\pm$0.027 &  0.28$\pm$1.05 &  0.83 & \\  
  13519 & $-$6.072$\pm$0.133 & 2.660$\pm$0.116 &   0.73 &  $-$0.743$\pm$0.145 &  0.484$\pm$0.108 &  0.886$\pm$0.068  & 0.13$\pm$0.14 &  0.05 & \\
  13804 & $-$5.444$\pm$0.132 & 0.871$\pm$0.114 &  17.03 &   0.492$\pm$0.139 & $-$0.955$\pm$0.113 &  1.074$\pm$0.059 &  0.57$\pm$2.77 &  0.95 & \\ 
  13860 & $-$4.677$\pm$0.080 & 2.105$\pm$0.075 &  12.67 &   0.759$\pm$0.083 &  0.473$\pm$0.083 &  0.894$\pm$0.042  & 0.02$\pm$1.61 &  0.85 & \\ 
  13908 & $-$4.712$\pm$0.062 & 0.990$\pm$0.056 &  15.58 &   1.130$\pm$0.069 & $-$0.578$\pm$0.062 &  1.270$\pm$0.034 &  0.79$\pm$1.06 &  0.74 & \\ 
  13918 & $-$4.584$\pm$0.060 & 0.646$\pm$0.053 &  17.56 &   1.375$\pm$0.067 & $-$0.852$\pm$0.059 &  1.618$\pm$0.032 &  0.30$\pm$0.92 &  0.65 & \\ 
  13931 & $-$4.050$\pm$0.060 & 1.459$\pm$0.054 &  15.81 &   1.577$\pm$0.067 &  0.099$\pm$0.061 &  1.580$\pm$0.033 &  0.25$\pm$0.65 &  0.60 & a \\ 
  14172 & $-$4.924$\pm$0.035 & 1.925$\pm$0.036 &  30.59 &   0.593$\pm$0.043 &  0.216$\pm$0.047 &  0.631$\pm$0.022 &  0.02$\pm$2.87 &  2.91 & \\  
\hline
\end{tabular}
a: this is 2MASS~J11171292-6120085 \citep{Gvaramadze2013}
}
\label{tab:ejections}
\end{table*}

\begin{table*}
\caption{Proper motions and related quantities for stars with impact parameters between 1 and 6 arcmin, and high relative PM ($> 0.6$ mas yr$^{-1}$).  The columns are as in Table~\ref{tab:ejections}.  Note that a timescale is only given in the final column if the uncertainty on the impact parameter leaves open the possibility that the star may have originated inside a radius of 1 arcmin around the cluster centre.}
{\centering
\begin{tabular}{lccrrrcccl}
\hline  
MS-II & \multicolumn{2}{c}{Proper motion} & Radius & \multicolumn{2}{c}{Relative PM, $PM_r$} & $|PM_r|$ & $IP$ & Travel time & Note \\
\# & \multicolumn{2}{c}{$\mu_{\alpha,*}$, $\mu_{\delta}$ mas/yr} & arcmin & \multicolumn{2}{c}{$\Delta\mu_{\ell,*}, \Delta\mu_{b}$ mas/yr} & mas/yr & arcmin & Myr & \\
\hline
 13280 & $-$6.309$\pm$0.072 & 3.156$\pm$0.071 & 16.09 & $-$1.144$\pm$0.076 & 0.860$\pm$0.078 & 1.431$\pm$0.038 & 2.06$\pm$1.20 & 0.67 & \\
 13377 & $-$5.636$\pm$0.042 & 2.736$\pm$0.036 & 24.36 & $-$0.364$\pm$0.050 & 0.713$\pm$0.046 & 0.801$\pm$0.023 & 3.48$\pm$1.97 & & \\
 13452 & $-$5.359$\pm$0.036 & 2.870$\pm$0.033 & 43.88 & $-$0.155$\pm$0.045 & 0.939$\pm$0.044 & 0.951$\pm$0.022 & 3.25$\pm$2.38 & 2.77 & \\
 13573 & $-$5.384$\pm$0.119 & 1.324$\pm$0.095 & 2.60 & 0.383$\pm$0.126 & $-$0.511$\pm$0.094 & 0.639$\pm$0.053 & 1.81$\pm$0.45 & & \\
 13708 & $-$6.518$\pm$0.130 & 1.093$\pm$0.120 & 2.70 & $-$0.590$\pm$0.129 & $-$1.138$\pm$0.127 & 1.282$\pm$0.064 & 2.59$\pm$0.10 & & b \\
 13766 & $-$6.460$\pm$0.068 & 1.986$\pm$0.065 & 5.52 & $-$0.860$\pm$0.072 & $-$0.285$\pm$0.072 & 0.906$\pm$0.036 & 3.12$\pm$0.46 & & \\
 14359 & $-$6.377$\pm$0.044 & 1.890$\pm$0.040 & 40.68 & $-$0.748$\pm$0.051 & $-$0.344$\pm$0.050 & 0.823$\pm$0.026 & 3.69$\pm$3.28 &  & c \\
\hline
\end{tabular}
b: this star narrowly missed exclusion on account of large parallax -- see text\\
c: the relative proper motion of this star is directed {\em towards} the cluster
}
\label{tab:near-fast}
\end{table*}

\begin{figure*}
\begin{center}
\includegraphics[width=1.35\columnwidth]{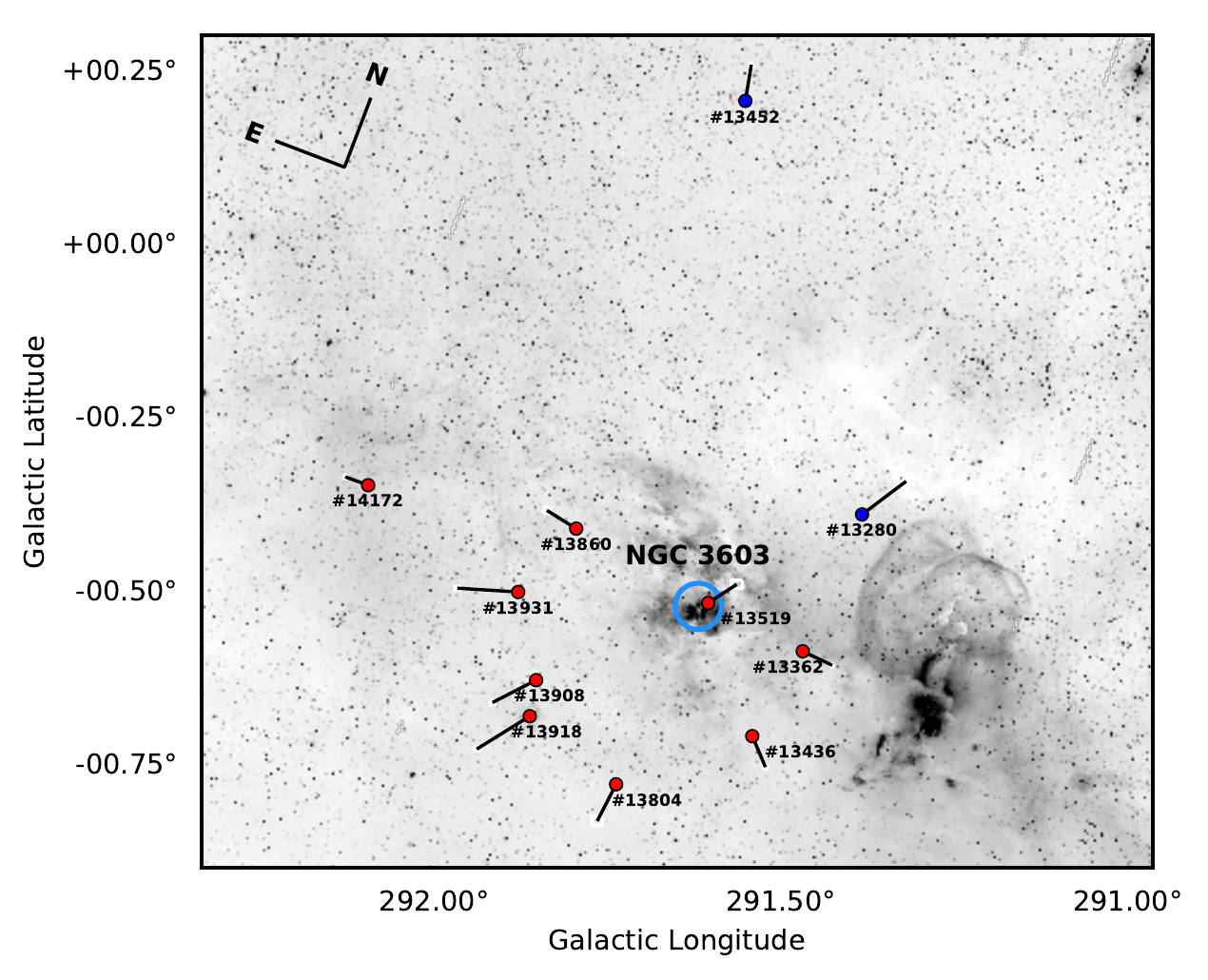}
\caption{
The on-sky distribution of the O-star candidates, with relative proper motion magnitude exceeding 0.6 mas/yr.  Objects coloured in red are those with impact parameters that are $<$ 1 arcmin (Table~\ref{tab:ejections}).  The two objects in blue are from Table~\ref{tab:near-fast}: they have impact parameter errors that allow a trajectory consistent with ejection at reduced probability.   
}
\label{fig:pms-onsky}
\end{center}
\end{figure*}

For the present purpose of initial classification, we note that at a distance of 7 kpc, a proper motion magnitude of 0.6 mas yr$^{-1}$ corresponds to an in-sky or tangential speed of 20 km~s$^{-1}$.  We will regard any object that exceeds this threshold as meeting the first of two criteria required of candidate ejections from NGC 3603. 38 objects qualify in this regard. The second criterion to be satisfied is that the trajectory of motion needs to have an impact parameter (closest distance of approach to NGC 3603 centre) under 1 arcmin, and the sense of motion needs to be {\em away} from the cluster.

Figure~\ref{fig:imppar} shows how the stars with impact parameter less than 15 arcmin break down according to magnitude of proper motion.  Among the 199 stars plotted there is a clear peaking of impact parameter to smaller values (irrespective of relative PM magnitude).  This pattern persists as the total list is reduced by cutting on an increasing minimum relative PM magnitude.  In the most extreme group, with $|PM_r| > 0.6$ mas yr$^{-1}$, 9 out the 38 stars have an impact parameter of under 1 arcmin.  These are our candidate ejections meeting the first two criteria.  They are coloured lighter pink in Figure~\ref{fig:relpms}, while their on-sky distribution is shown in \ref{fig:pms-onsky}.  Important properties for this set of objects are given in Table~\ref{tab:ejections}.  

Of the 9 candidates, 1 object (VPHAS-OB1-13519, or \#13519 in Table~\ref{tab:ejections}) is a relevant inclusion in that $|PM_r| = 0.886\pm0.068$ mas~yr$^{-1}$ is well above threshold for the first criterion -- otherwise, it is located within 1 arcmin of cluster centre, and so must have an impact parameter $<1$~arcmin.  The direction of its $PM_r$ is almost opposite to that it would have if merely a foreground contaminant.   \#13519 could be a very early-stage ejection, exiting the central region at a tangential speed of $\sim30$ km s$^{-1}$ (at 7 kpc)

More interest attaches to the group of 7 separated from the cluster centre by between 9.68 and 17.56 arcmin.  Representative values for the impact parameter and error in this group are respectively $\sim$0.3 and $\sim$1~arcmin (see Table~\ref{tab:ejections}) -- implying that our second criterion may actually scoop up stars originating from within a radius of $\sim1.3$ arcmin.  At the shortest radius in this group, the probability of a star having the right direction of travel to satisfy the second criterion (if all directions of travel are equally likely) is 0.043.  The probability that it also has $|PM_r| > 0.6$ is empirically $38/288 \simeq 0.13$. Hence the total combined chance is $\sim$0.006. For the most far-flung member of the group, at almost twice the radius, the probability essentially halves. This general level of individual probability hints that perhaps one of the seven objects could be a false positive (given the total population drawn from).

One of the group of 7 has already been identified as a likely ejection by \cite{Gvaramadze2013}: it is 2MASS~J11171292-6120085, an O6V star with an associated suitably-offset bow shock.  In MS-II it is VPHAS-OB1-13931 (\#13931 in Table~\ref{tab:ejections}).  It is located nearly 16 arcmin from the centre of NGC 3603 and its relative PM is 1.58 mas yr$^{-1}$ (or $\sim$52 km s$^{-1}$ at 7 kpc).  It is the object almost directly to the left of cluster centre in Figure~\ref{fig:pms-onsky}. However, its proposed partner object, WR 42e, displaced to the opposite side of the core of NGC 3603 is not supported as an ejection: its $|PM_r|$ is small at 0.125$\pm$0.034 mas/yr.  Similarly, for none of the three additional ejection candidates, RFS 1, 2 and 8, put forward by \cite{RomanLopes2016} is their relative proper motion significant (the largest is $0.178\pm 0.020$ mas yr$^{-1}$, obtained for RFS 1).  In the case of RFS 8, some 29 arcmin away from the cluster centre, runaway status is entirely ruled out as the implied timescale since ejection would have to exceed 10 Myr.  RFS 1 and 2 are respectively only 0.7 and 1.0 arcmin away from cluster centre, leaving open the possibility of ejection should spectroscopic observations reveal significant relative radial velocity.  

.

The on-sky pattern traced by the group of 7 is strikingly a half ring to the south of the core region.  The one object, \#14172, sitting outside this zone also distinguishes itself in Table~\ref{tab:ejections} as the only candidate with an estimated time of flight appreciably larger than 1 Myr.  Otherwise, the clear norm is a flight time of under 1 Myr.  In section~\ref{sec:discussion} we will consider what this tidy pattern of ejections may signify.


\subsection{Other high relative proper motion objects in the region}
\label{sec:other-highpm}

Figure~\ref{fig:imppar} shows that there are high relative PM stars in the sample with trajectories that do not pass close to the cluster centre.  These will be the stars shown in light pink and magenta, at impact parameters exceeding 1 arcmin.  In particular there is a group of fast ($|PM_r| > 0.6$ mas yr$^{-1}$) 'near misses' with impact parameters up to 4 arcmin, which are worth brief consideration (their properties are in Table~\ref{tab:near-fast}).  The uncertainties on the impact parameters of 2 of them are large enough that it cannot be ruled out that they may have been ejected from within 1 arcmin of the cluster centre.  For this reason they have been included in Figure~\ref{fig:pms-onsky} and coloured in blue.  One of the two, \#13280, may continue the ring of ejections discussed above in section~\ref{sec:ejections}.  The other, \#13452, represents a contrast in that it is much further away from NGC 3603 and would have to have been ejected about 3 million years ago if indeed it was ejected. 

The status of the remaining 5 objects in Table~\ref{tab:near-fast}, that appear never to have been in the cluster core, is less certain.  There is a case to be made that one of them, \#13708, is in fact a foreground star in that its Gaia DR2 parallax is $0.3297\pm0.0731$, a $4.5\sigma$ measurement. It has only remained in the sample because the $5\sigma$ limit is not breached -- indeed it is the object with the highest measured parallax retained and it is responsible for the lowest occupied histogram bin in Figure~\ref{fig:edsd-distance}.

We can try a combination of high relative PM {\em and} high impact parameter as the means to identify 'contaminant' objects.  We view it as improbable that higher relative PM stars ($|PM_r| > 0.42$ mas yr$^{-1}$, or $> 2\sigma$ in the 2-D dispersion), with impact parameters larger than some minimum value are associated with the cluster.  Anticipating the result of the next section, we look for impact parameters exceeding 6 arcmin.  Such a cut pulls out 30 objects.  It is interesting to note that 20 in this group have significantly negative longitudinal $|PM_r|$ ($< -0.4$ mas yr$^{-1}$) -- a property consistent with being in the foreground to the cluster.  Indeed, these objects dominate the negative 'shoulder' seen in Figure~\ref{fig:relpms} (lower panel), and all are at an angular separation of at least 20 arcmin from the centre of NGC 3603.  The remaining 10 stars are -- with one exception only -- over $\sim$30 arcmin distant.  The exception is \#13390 that is at a radius of 9.1 arcmin, with an estimated impact parameter of 6.5 arcmin: it is an IR-bright object ($K = 9.71$) that also happens to be highly obscured ($A_0 = 9.84$ mag).  Potentially, it is in the background to the cluster.

\section{The O star hinterland}
\label{sec:size}

The size of region on the sky that should be associated with NGC 3603 is  challenging to define.  Based on $K_s$ photometry, \cite{Nurnberger2002} constructed a stellar surface density map to a range of limiting magnitudes and concluded that at a radius of 2.5$\pm0.25$ arcmin a mean field takes over from a declining cluster density profile. A King model was fit to the data, in which the core and tidal radii were respectively 23 and 1300 arcsec.  This has been revisited by \cite{Harayama2008}, using NIR adaptive-optics data better able to resolve the core. They revised the core radius down to 4.8 arcsec,  and proposed a tidal radius of 1260 arcsec (21 arcmin) on general dynamical grounds.  The latter is much the same as the \cite{Nurnberger2002} estimate.  Importantly, \cite{Harayama2008} elaborated the evidence in favour of significant mass segregation such that the bright core contains a relative concentration of the most massive (O) stars -- a result that was later reinforced by \cite{Pang2013}.     

\begin{figure*}
\begin{center}
\includegraphics[width=0.95\columnwidth]{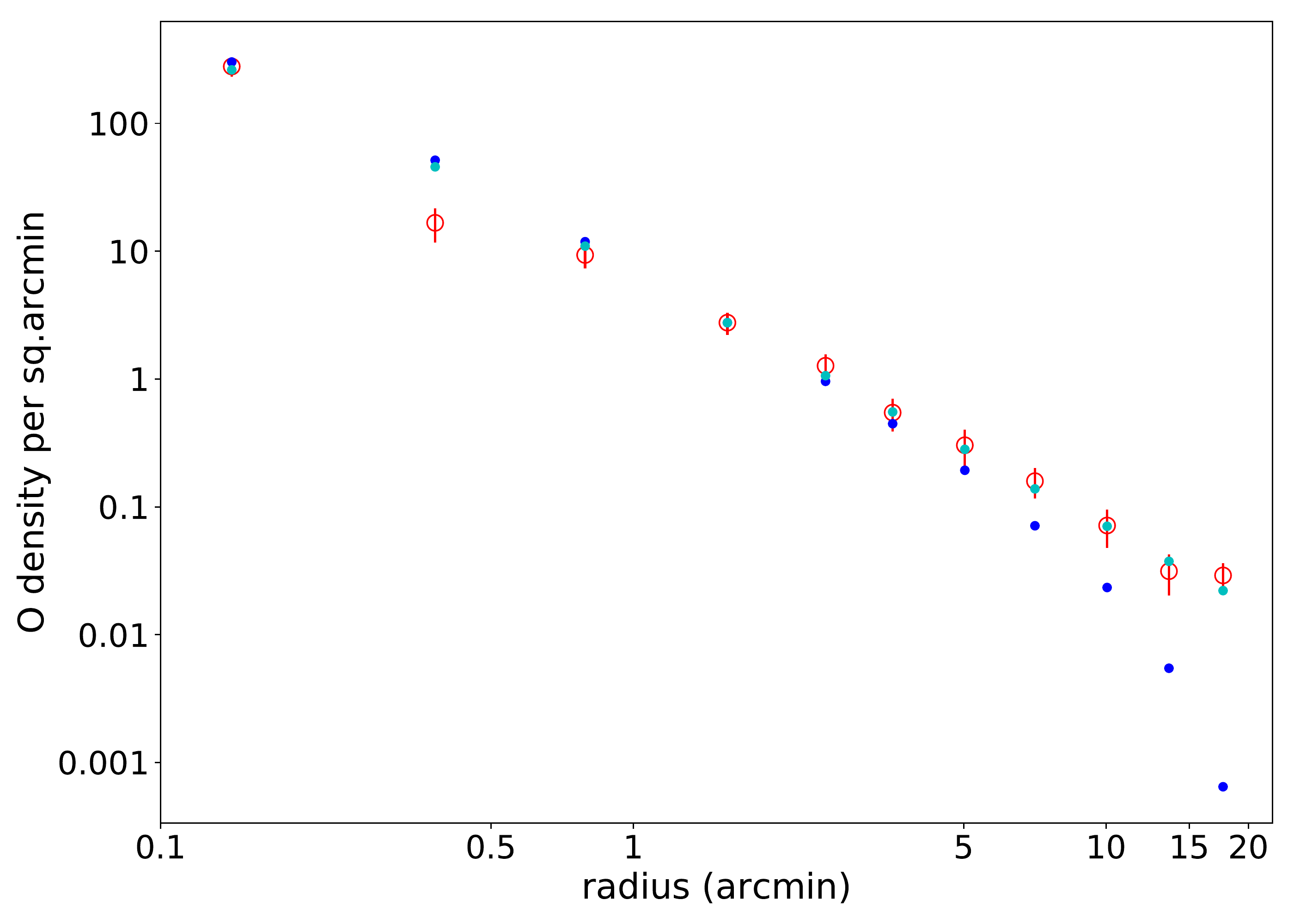}
\includegraphics[width=0.95\columnwidth]{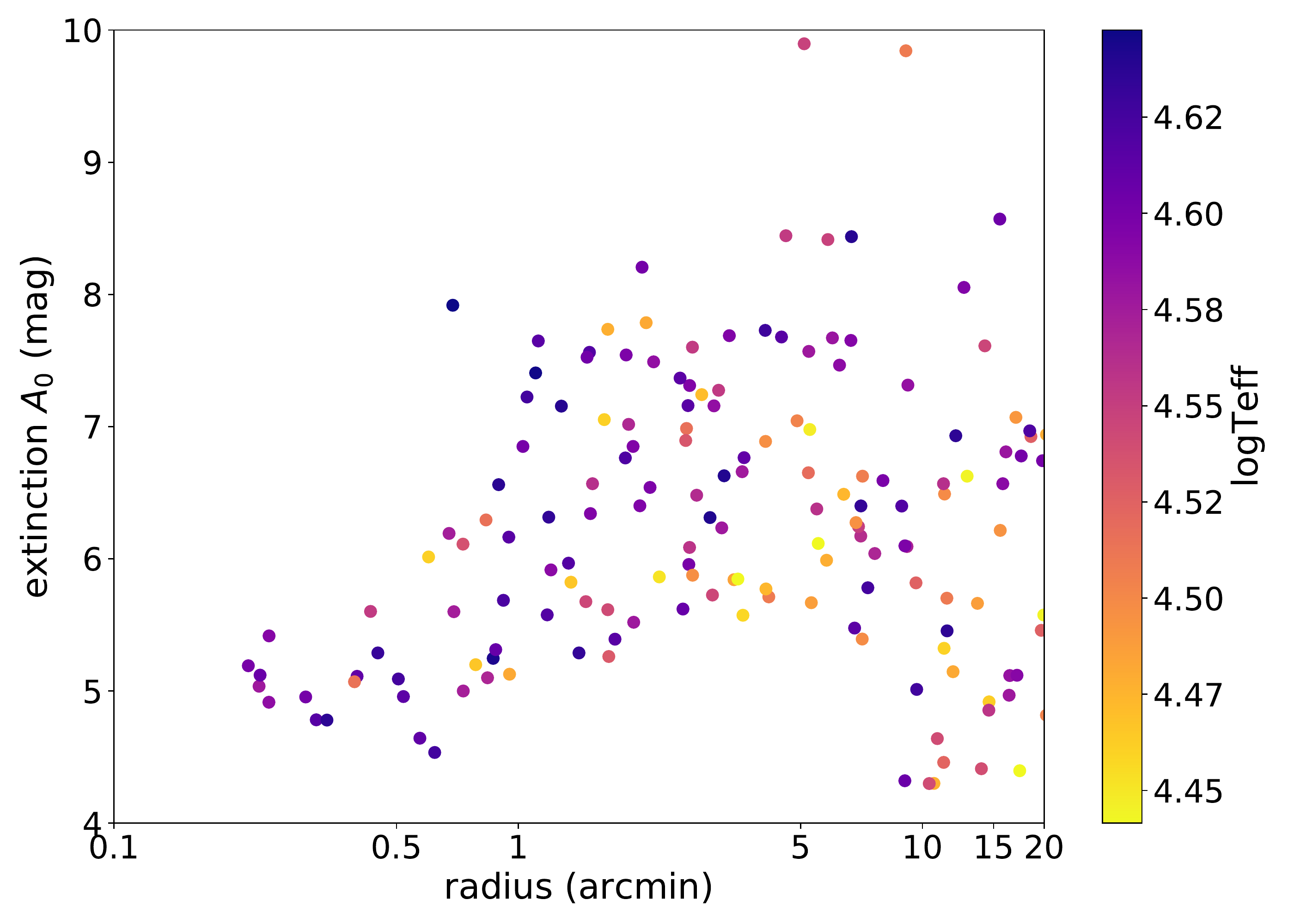}
\caption{On a log-log scale, the upper panel shows the density of selected O stars as a function of radius out to 20 arcmin (red circles). The error bars are Poissonian. The O star count inside 0.2 arcmin is equated to the number visible in Figure 2 of Melena et al (2008): it is likely to be a lower limit.  The blue data points trace the King profile obtained by Harayama et al (2008), while the cyan points show the same profile computed for a very large tidal radius.  Both model curves have been rescaled to match the O star density in the 1.0-2.0 arcmin bin.  The lower panel shows the radial extinction distribution for comparison,  with the data points coloured according to the MS-II effective temperature estimate.
}
\label{fig:density}
\end{center}
\end{figure*} 

Because we have a honed selection of O stars, across a wide field, we have the opportunity to review the radial density profile, restricted to the context of the most massive stars for which the enveloping 'field' density would most likely be low.  

We have determined the density profile out to a maximum radius of 20 arcmin, supplementing the Gaia DR2 cross-matched list with 17 objects in our initial selection that so far lack matches -- this adds, as a crude average, one object per bin.  We leave out \#13390 and \#13708 (see section~\ref{sec:other-highpm}).  The resultant profile is shown in the upper panel of Figure~\ref{fig:density}. 
As a comparison, the King model proposed by \cite{Harayama2008} is shown rescaled to coincide with the measured density in the radius range 1.0 -- 2.0 arcmin.  
Two features stand out.  First, the model rolls off a little quickly relative to the observed O stars, possibly indicating too short a tidal radius.  If instead the tidal radius is set to be very large (cyan points in the plot), the match is better but high given that some contaminating objects almost certainly remain present.  Second, over the radius range 0.2 to 1 arcminute, the King-model trend is relatively underpopulated.  Given that the count of O stars inside 0.2 arcmin is most likely an underestimate \citep[see][]{Melena2008}, this could be a reflection of the mass segregation described by both \cite{Harayama2008} and \cite{Pang2013} that is most evident inside a radius of 0.5 - 1 arcmin.  

The King model superposed in Figure~\ref{fig:density} works quite well with minimal contamination out to a radius of $\sim$ 5 arcmin.  At larger radii it becomes necessary to consider a sliding scale of options ranging from a lot of presumed field contamination combined with the fall-off predicted by Harayama et al.'s (2008) King model, through to a little contamination on top of a distribution subject to a larger a tidal limit.  If the tidal radius is $\sim 21$ arcmin, then the O star count predicted by the rescaled King model, outside a radius of 0.2 arcmin, is $\sim$138 -- we count 100 to 5 arcmin, 132 to 10 arcmin, and 166 to 20 arcmin.   

The lower panel of Figure~\ref{fig:density} provides some corroborating insight into what is going on -- essentially, the dependence of O-star extinction on radius is orderly and in keeping with the earlier mapping to $\sim$4 arcmin of colour excess by \cite{Sung2004}. But from around 6 arcmin, outwards, the order begins to break down and a number of O stars begin to present with lower extinction relative to the trend seen at shorter radii.  This could well be foreground contamination revealing itself.  

We conclude that the 'halo' of NGC 3603 may very well extend beyond 5 arcmin, and that essentially all the O stars drawn from the MS-II catalogue inside this radius are associated objects.  Out of the 100 objects inside 5 arcmin, just 15 were mentioned in the literature prior to MS-II (their names are matched to the MS-II and Gaia DR2 identifiers in table~\ref{tab:names}). The lower panel of Figure~\ref{fig:density} demonstrates that most are expected to have effective temperatures of $\sim$35 kK or more -- implying masses exceeding $\sim$20~M$_{\odot}$ \citep[see][]{Ekstrom2012}.



\section{Discussion}
\label{sec:discussion}

\subsection{The pattern of O-star ejections}

Our selection and survey of O-star proper motions relative to NGC 3603 has revealed at 9 credible ejections (Table~\ref{tab:ejections}) and 2 further less certain examples (Table~\ref{tab:near-fast}).   The scale of the measured proper motions for all but two of the candidates indicates a time since ejection of under a million years.  This is entirely congruent with findings that the bright compact centre of the cluster, contained inside a radius of 1 arcmin, is no older than 1--2 Myr \citep[e.g.][]{Sung2004,Kudryavtseva2012}.  Indeed our result endorses this young age, and carries no dependence on the still uncertain distance to NGC 3603.  

The two objects with trace-back times closer to 3 Myrs (\#14172 and, with less precision, \#13452) may be evidence of an earlier phase of star-forming activity.  Or they might not survive further scrutiny.  We note now that their extinctions are relatively low at respectively $A_0 = 4.31$ and 4.40 -- to be compared with $5 < A_0 > 8$ for all the ejections in the last 1 Myr.   To pursue this question further requires expanding the search area to pick up more fast-moving ejections (\#13452 is right on the edge of our $1.5\times1.5$ sq.deg region), and/or follow-up spectroscopy to better characterize these objects.  

The most remarkable feature of the on-sky arrangement of the candidate ejections is that the 7 most convincing are arranged in a semi-oval emphasising the south side of the cluster core (Figure~\ref{fig:pms-onsky}). At most, one of these might be a false positive. An eighth (\#13280) helps to spread the pattern a bit more into the north but it is more marginal for inclusion since its impact parameter is a factor of a few larger.  Statistically there are no grounds to suspect that a bias in the sample as a whole, favouring the south over the north, shapes this -- the main difference is that north of cluster centre the density of objects drops away more quickly than towards the south.  There is no reason to suspect a north-south extinction bias either \citep[see the extinction maps of][]{Marshall2006,PlanckXI}.  The simplest option is to accept, provisionally, that the pattern reflects a reality beyond mere coincidence.

The pattern seen is in striking contrast to the also-orderly pattern of ejections from Westerlund 2 \citep{Drew2018}: in that case a distinctly linear pattern was present with ejections located on either side of the cluster.  It was argued that sub-cluster merging to produce present-day Westerlund 2 could be responsible for the preferred axis of the ejections.  Could something similar be involved here?  

\cite{Fukui2014} have presented detailed CO observations in the vicinity of NGC 3603, and have deduced from them evidence of a cloud-cloud collision that took place around a million years ago.  And they made a link between NGC 3603 and Westerlund 2 as two examples of 'super star clusters' -- suggesting that this status owes something to cloud-cloud collisions.  In this context we note that the most massive molecular cloud component \citep[reported by][]{Fukui2014}, implicated in the case of NGC 3603, is centred on $\ell = 291^{\circ}.58$, $b = -0^{\circ}.42$.  This is 5--6 arcmin north of the core of NGC 3603 and, in the plane of the sky, on the opposite side to the semi-circle of ejections. If this cloud carried most of the momentum in the putative collision, we might anticipate that most of the ejections would appear on the same side (contrary to what is seen).  

From a theoretical perspective, the formation of NGC 3603 has been discussed in terms of a monolithic process -- either as a single intense star-forming event \citep{Banerjee2014} or via the prompt assembly of a compact group of sub-clusters \citep{Banerjee2015}. 
The former is favoured by \citep{Banerjee2015} and resembles the early assembly concept discussed by \cite{Fujii2013}. Dynamical stellar ejection is the only relevant ejection process here, for the reason that the youth of NGC 3603 should mean no supernovae have exploded yet. In this context, there need be no particular expectation of a particular pattern of ejections: random vectors, but on a timescale comparable with the formation event would seem plausible. The times since ejection for all objects making up the ring fall within a quite narrow range, from 0.60 to 0.95 Myr.  Expressed as a mean and standard deviation, the ring could be linked to an event 0.75$\pm$0.11 Myr ago.  However the propagated errors on the individual timescales do not exceed $\sim$0.05 Myr (if we assume the point of origin within the cluster for each ejection is unknown to within 0.2 arcmin). Hence, a modest spread in time of ejection of up to $\sim$200,000 years appears more likely. To make progress, radial velocities for the candidate ejections would be a good next step as this would build a better view of the full three-dimensional geometry.  It is possible that the projection into two dimensions exaggerates the degree of coherent spatial organisation present.  

A last point of interest is that Figure~\ref{fig:pms-onsky} and the data in Tables~\ref{tab:ejections} and \ref{tab:near-fast} indicate that there are potentially two 'pairings' of ejections: \#13280 may pair with either  \#13908 and \#13918, while \#13362 and \#13860 are moving in close to opposite directions.  This leaves 3 (or 4) of the stars in the ring of ejections without evident partners -- one of these is \#13931 for which a partner has been claimed \citep{RomanLopes2016}, but has only a small proper motion.   

\subsection{The significance of the O star halo}

We noted in the introduction that much of the work to date on NGC 3603 has focused on the inner $\sim$arcminute around the brilliant cluster core.  In this study, encouraged by the initial findings reported by MS-II, we have expanded the area examined up to over a square degree.  With the support of Gaia DR2 proper motions, the case has been built for an extensive hinterland of associated O stars that persists to at least a radius of 5 arcmin.  The distribution is consistent with Harayama et al's (2008) preferred King model, for which a tidal radius of 21 arcmin was adopted.  \cite{Harayama2008} pointed out that the tidal radius is hardly constrained at all and could be even larger.  The clear implication is that a number of the O stars in our sample between $\sim$5 and $\sim$20 arcmin could be halo members too.  It will take full space motions to clarify this.

It has been mooted before that NGC 3603 might be a Galactic counterpart to R136 in the 30 Doradus region of the LMC.  In pursuit of this point, \cite{Melena2008} argued that R 136, the central cluster of 30 Doradus, contains between 1.1 and 2.4 times as many very high mass stars ($M_{{\rm bol}} < -10$) as found in the core of NGC 3603 -- that is, the scale factor is not so large.

We now make essentially the same comparison, basing it on the wider environment rather than the core region. \cite{Evans2011} list 100 stars with $K_S < 15.5$ ($M_K < -3$ or $M_{{\rm bol}} < -6.8$) in the annulus between 0.2 to 1 arcmin, reaching out into R 136's 'halo'. This can be compared with the count here to the same $M_K$ limit, in the equivalent angular range, after rescaling for the much shorter distance to NGC 3603 of $7\pm 1$ kpc (with R136/30 Dor at a distance of 50~kpc, the rescale is 7$\times$, giving an angular radius range of 1.4 to 7 arcmin).  The analogous count is $\sim$40, implying a scale-down from R136/30 Dor by about a factor of $\sim$2.5. This is at the upper end of the \cite{Melena2008} range, and is just about compatible with it.  But we have to differ with their conclusion \citep[and that of][]{Moffat2002} that there is "no surrounding massive halo of cluster stars".  There evidently is.  The difference between then and now is the availability of calibrated wide field multi-colour photometry.

\section{Conclusions} 
\label{sec:conclusions}

The crossmatch we have carried out of 288 stars in the hinterland of the massive young cluster, NGC 3603, of  high-purity O-star candidates in MS-II and the Gaia DR2 release has had two main outcomes.

1.  Our appraisal of the relative proper motions has revealed up to 11 candidate O star ejections.  Nine of these have been ejected within the last one million years.  Indeed the timescale spread is limited to 0.6--0.95 Myrs for eight of them.  This lends clear support and an interesting datum to earlier photometric studies that have argued the central cluster is no more than 1--2 Myrs old \citep{Sung2004, Melena2008, Kudryavtseva2012}.  The on-sky pattern of these ejections comes as a surprise in that 7 are arranged in a partial ring of radii spanning 9--18 arcmin, favouring the south.  Radial velocities are needed to begin to add the third dimension.  It is hard to see how this pattern would arise from a cloud-cloud collision, given what we know about the placement of molecular clouds in the area \citep{Fukui2014}.  In this respect NGC 3603 is different from Westerlund 2, where the O-star ejections are aligned with the axis of a putative prior cloud-cloud collision \citep{Drew2018}.  It seems more likely that the ejections from NGC 3603 have arisen from a first cluster core collapse \citep[see e.g.][]{Fujii2013, Banerjee2014}.

2.  We have put forward evidence of a notable halo of O stars around the central cluster of NC 3603 reaching out to a radius of at least 5 arcmins.  At larger radii, some non-member contamination is very likely.  We have counted of order 100 O stars in the halo based on a comparison with the King model obtained by \cite{Harayama2008}.   Earlier work doubted that such a halo exists.  Now we have detected it, it can be estimated that the HD~97950 cluster at the heart of NGC 3603 is part of a larger entity that parallels the core-halo structure of the R136/30 Dor region in the Large Magellanic Cloud, albeit scaled down to around 40\% of the total O-star population.  

Appendix B (online only) provides the derived relative proper motion data forming the basis of this study, as table~\ref{tab:crossref}.

\section*{Acknowledgements}

Use has been made of data products from the Two Micron All Sky Survey, which is a joint project of the University of Massachusetts and the Infrared Processing and Analysis Center/California Institute of Technology, funded by the National Aeronautics and Space Administration and the National Science Foundation. This work has also used data from the European Space Agency mission Gaia (https://www.cosmos.esa.int/gaia), processed by the Gaia Data Processing and Analysis Consortium (DPAC, https://www.cosmos.esa.int/web/gaia/dpac/consortium). Funding for the DPAC has been provided by national institutions, in particular the institutions participating in the Gaia Multilateral Agreement.

Much of the analysis presented has been carried out via TopCat \citep{Taylor2005}.


JED and MM acknowledge the support of a research grant funded by the Science, Technology and Facilities Council of the UK (STFC, ref. ST/M001008/1). NJW acknowledges receipt of an STFC Ernest Rutherford Fellowship (ref. ST/M005569/1).  We thank the referee of this paper for helpful comments.


\bibliographystyle{mnras}
\bibliography{JEDpapers}

\appendix

\section{Objects in the literature prior to MS-II}

Table~\ref{tab:names} identifies objects in the sample of 288 stars that were known before the compilation of the MS-II catalogue and Gaia DR2 release.  The obects' prior names appear in the middle column of the table and are given in the form used by SIMBAD. The entries are in order of MS-II VPHAS-OB1-nnnnn catalogue number, which in turn is ordered by Galactic longitude.  The unique identifier for the Gaia DR2 cross-match is also listed.  

\begin{table}
\caption{Cross-matched names}
{\centering
\begin{tabular}{lll}
\hline
MS-II  &  Prior names  & Gaia DR2 \\
list \# &            & identifier      \\ 
\hline
13089 & WR 42a               & 5337240002035340288 \\
13471 & WR 42e, RFS 5 and    & 5337420974779493248 \\
      & 2MASS J11144550-6115001 &                  \\
13529 & Cl* NGC 3603 MDS 76  & 5337418019842038912 \\
13546 & Cl* NGC 3603 MDS 7   & 5337418015513333888 \\
13555 & Cl* NGC 3603 MDS 3   & 5337418019842060928 \\
13570 & Cl* NGC 3603 Sher 23 & 5337418397799183872 \\
13572 & RFS 1, MTT 31        & 5337417985482256768 \\
13576 & Cl* NGC 3603 Sher 18 & 5337418015513337472 \\
13579 & Cl* NGC 3603 Sher 22 & 5337418019842057088 \\
13581 & Cl* NGC 3603 Sher 47 & 5337418015513340160 \\
13584 & Cl* NGC 3603 MTT 47  & 5337418191640752896 \\
13589 & RFS 2, MTT 58        & 5337417985482248576 \\
13594 & Cl* NGC 3603 MTT 25  & 5337417813683617152 \\
13931 & 2MASS J11171292-6120085 & 5337403073354793088 \\
13954 & RFS 8                   & 5337401724739725184 \\
\hline
\end{tabular}
 }
\label{tab:names} 
\end{table}

\section{The complete cross-matched sample}

Table~\ref{tab:crossref} lists the MS-II and Gaia DR2 identifiers along with sky positions, $\log T_{eff}$, extinction $A_0$ (mag) from MS-II of all 288 stars making up the main sample discussed in this paper.  For convenience, 2MASS $K$ magnitude is also cited (column 6).  The radius in column 7 is the angular distance from the adopted cluster centre, at Galactic coordinates, $\ell = 291{\circ}.617$, $b = -0^{\circ}523$.  Later columns present the derived relative proper motions, $PM_r$, and impact parameters, $IP$.

\onecolumn
{\centering

\begin{landscape}

\begin{longtable}{cccccccccccl}
\caption{Parameters of the full sample.}\\
 \hline
MS-II & \multicolumn{2}{c}{RA,Dec J2000} &$\log{T_{{\rm eff}}}$ & $A_0$ & $K$ & radius & DR2 Name & \multicolumn{2}{c}{$PM_r$} &$|PM_r|$ & \ \ \ \ \ $IP$  \\
list \# & hh:mm:ss & dd:mm:ss &  & mag & mag & arcmin &  &  $\mu_{\ell,*}$ mas/yr &  $\mu_{b}$ mas/yr & mas/yr & \ \ \ arcmin  \\
\hline
\hline
12469 & 11:09:48.23 & -60:51:53.28 & 4.46 & 5.33 & 10.72 & 45.26 & 5337257555529491456 & 0.33 $\pm$ 0.05 & -0.14 $\pm$ 0.05 & 0.36 $\pm$ 0.03 & 10.54 $\pm$ 8.07  \\
12476 & 11:09:44.95 & -60:53:22.19 & 4.53 & 4.93 & 11.45 & 44.83 & 5337257422423768064 & -0.33 $\pm$ 0.05 & 0.19 $\pm$ 0.05 & 0.38 $\pm$ 0.03 & 16.49 $\pm$ 7.43  \\
12498 & 11:09:43.62 & -60:55:35.14 & 4.56 & 6.34 & 9.62 & 43.90 & 5337255871902117120 & 0.01 $\pm$ 0.06 & -0.14 $\pm$ 0.06 & 0.14 $\pm$ 0.03 & 43.34 $\pm$ 3.18  \\
12499 & 11:07:31.43 & -61:33:38.67 & 4.54 & 3.71 & 12.03 & 57.37 & 5337165063445617280 & -0.29 $\pm$ 0.05 & 0.11 $\pm$ 0.05 & 0.31 $\pm$ 0.02 & 50.20 $\pm$ 5.35  \\
12503 & 11:09:28.79 & -61:00:08.74 & 4.47 & 4.18 & 11.65 & 43.63 & 5337296248933283328 & -0.55 $\pm$ 0.06 & -0.11 $\pm$ 0.06 & 0.56 $\pm$ 0.03 & 7.35 $\pm$ 5.11  \\
12517 & 11:08:31.01 & -61:17:58.25 & 4.53 & 5.85 & 12.20 & 47.60 & 5337267455469411584 & -0.44 $\pm$ 0.07 & 0.10 $\pm$ 0.07 & 0.45 $\pm$ 0.03 & 29.08 $\pm$ 6.83  \\
12550 & 11:09:11.90 & -61:09:13.80 & 4.50 & 3.83 & 12.54 & 43.21 & 5337293465794355072 & 0.22 $\pm$ 0.04 & 0.52 $\pm$ 0.05 & 0.57 $\pm$ 0.02 & 34.80 $\pm$ 2.69  \\
12561 & 11:08:01.33 & -61:31:10.69 & 4.47 & 3.43 & 12.91 & 53.25 & 5337165407043165440 & -0.30 $\pm$ 0.05 & 0.04 $\pm$ 0.05 & 0.31 $\pm$ 0.02 & 39.09 $\pm$ 6.48  \\
12564 & 11:09:37.18 & -61:03:57.00 & 4.53 & 3.99 & 11.45 & 41.42 & 5337248690716344704 & -0.21 $\pm$ 0.04 & 0.24 $\pm$ 0.05 & 0.32 $\pm$ 0.02 & 33.77 $\pm$ 4.70  \\
12584 & 11:10:27.66 & -60:50:32.58 & 4.46 & 5.68 & 10.89 & 42.07 & 5337632355894016896 & -0.21 $\pm$ 0.05 & 0.10 $\pm$ 0.05 & 0.23 $\pm$ 0.03 & 7.91 $\pm$ 12.00  \\
12589 & 11:10:30.58 & -60:50:02.55 & 4.52 & 5.33 & 10.65 & 42.10 & 5337632458973245952 & -0.41 $\pm$ 0.04 & 0.17 $\pm$ 0.05 & 0.44 $\pm$ 0.02 & 5.19 $\pm$ 5.59  \\
12594 & 11:11:25.30 & -60:34:04.61 & 4.63 & 7.67 & 12.39 & 49.51 & 5337640567872208512 & -1.05 $\pm$ 0.15 & 0.19 $\pm$ 0.12 & 1.07 $\pm$ 0.07 & 21.11 $\pm$ 5.93  \\
12615 & 11:09:14.33 & -61:14:39.87 & 4.55 & 5.96 & 12.74 & 42.39 & 5337222749113282432 & -0.33 $\pm$ 0.08 & 0.13 $\pm$ 0.08 & 0.36 $\pm$ 0.04 & 28.67 $\pm$ 9.46  \\
12641 & 11:10:03.40 & -61:03:07.50 & 4.61 & 4.02 & 11.63 & 38.67 & 5337248214013314048 & -0.26 $\pm$ 0.04 & 0.33 $\pm$ 0.04 & 0.41 $\pm$ 0.02 & 31.59 $\pm$ 3.26  \\
12650 & 11:10:13.89 & -61:01:46.21 & 4.51 & 4.58 & 12.36 & 37.96 & 5337254158248689280 & -0.17 $\pm$ 0.06 & 0.22 $\pm$ 0.05 & 0.28 $\pm$ 0.03 & 29.73 $\pm$ 6.68  \\
12659 & 11:09:12.69 & -61:20:34.63 & 4.46 & 2.94 & 11.45 & 42.79 & 5337220481370328320 & -0.20 $\pm$ 0.04 & -0.19 $\pm$ 0.06 & 0.27 $\pm$ 0.03 & 11.39 $\pm$ 10.74  \\
12661 & 11:09:58.27 & -61:07:55.21 & 4.49 & 4.91 & 12.14 & 37.95 & 5337247522488856832 & -0.26 $\pm$ 0.05 & 0.05 $\pm$ 0.05 & 0.27 $\pm$ 0.03 & 13.45 $\pm$ 8.24  \\
12693 & 11:10:33.58 & -61:01:29.61 & 4.52 & 5.05 & 10.71 & 35.87 & 5337254330047418880 & -0.38 $\pm$ 0.05 & -0.09 $\pm$ 0.04 & 0.39 $\pm$ 0.02 & 9.04 $\pm$ 4.40  \\
12706 & 11:10:38.75 & -61:00:44.42 & 4.47 & 4.62 & 12.22 & 35.61 & 5337254433126814080 & 0.03 $\pm$ 0.05 & -0.01 $\pm$ 0.05 & 0.03 $\pm$ 0.03 & 12.51 $\pm$ 31.68  \\
12710 & 11:09:22.36 & -61:23:16.87 & 4.50 & 3.73 & 11.54 & 42.02 & 5337219798508173440 & -0.04 $\pm$ 0.04 & 0.10 $\pm$ 0.04 & 0.11 $\pm$ 0.02 & 40.86 $\pm$ 4.70  \\
12724 & 11:12:02.02 & -60:38:32.77 & 4.53 & 5.59 & 11.75 & 43.34 & 5337639021684372608 & -0.11 $\pm$ 0.06 & -0.28 $\pm$ 0.05 & 0.30 $\pm$ 0.03 & 41.55 $\pm$ 2.88  \\
12729 & 11:10:56.54 & -60:58:30.28 & 4.47 & 6.11 & 12.19 & 34.73 & 5337253093096924288 & -0.19 $\pm$ 0.09 & 0.17 $\pm$ 0.08 & 0.25 $\pm$ 0.04 & 19.07 $\pm$ 13.60  \\
12733 & 11:08:28.75 & -61:41:33.18 & 4.47 & 5.65 & 11.07 & 54.12 & 5337115138740964352 & -0.25 $\pm$ 0.04 & 0.06 $\pm$ 0.05 & 0.26 $\pm$ 0.02 & 48.90 $\pm$ 5.15  \\
12736 & 11:09:27.91 & -61:24:57.75 & 4.47 & 2.18 & 11.95 & 41.69 & 5337218939514718976 & -0.87 $\pm$ 0.05 & -0.12 $\pm$ 0.06 & 0.88 $\pm$ 0.03 & 18.98 $\pm$ 2.60  \\
12754 & 11:08:28.87 & -61:43:39.84 & 4.48 & 5.22 & 12.07 & 55.13 & 5337114554625407232 & -0.64 $\pm$ 0.06 & 0.17 $\pm$ 0.06 & 0.66 $\pm$ 0.03 & 50.81 $\pm$ 2.34  \\
12784 & 11:10:05.83 & -61:17:26.65 & 4.63 & 4.53 & 11.60 & 36.20 & 5337221443447103744 & -0.04 $\pm$ 0.05 & 0.42 $\pm$ 0.05 & 0.42 $\pm$ 0.02 & 34.23 $\pm$ 1.50  \\
12800 & 11:11:43.24 & -60:49:56.33 & 4.59 & 4.12 & 11.02 & 35.59 & 5337630221251708160 & -0.10 $\pm$ 0.04 & -0.10 $\pm$ 0.04 & 0.14 $\pm$ 0.02 & 33.32 $\pm$ 5.02  \\
12803 & 11:10:30.99 & -61:11:25.96 & 4.61 & 4.36 & 9.66 & 33.45 & 5337244193924127744 & -0.14 $\pm$ 0.05 & -0.01 $\pm$ 0.05 & 0.14 $\pm$ 0.02 & 7.06 $\pm$ 12.08  \\
12807 & 11:11:55.61 & -60:46:28.71 & 4.48 & 4.50 & 12.59 & 37.23 & 5337636685221851776 & 0.06 $\pm$ 0.06 & -0.04 $\pm$ 0.05 & 0.07 $\pm$ 0.03 & 2.13 $\pm$ 32.71  \\
12809 & 11:11:55.30 & -60:46:38.42 & 4.56 & 4.43 & 10.93 & 37.13 & 5337636650862110080 & -0.18 $\pm$ 0.04 & -0.21 $\pm$ 0.04 & 0.28 $\pm$ 0.02 & 36.31 $\pm$ 1.52  \\
12813 & 11:11:09.25 & -61:00:50.84 & 4.57 & 7.41 & 12.28 & 32.25 & 5337252783859458048 & -0.09 $\pm$ 0.13 & 0.10 $\pm$ 0.11 & 0.14 $\pm$ 0.06 & 22.12 $\pm$ 22.27  \\
12816 & 11:10:22.46 & -61:14:45.52 & 4.52 & 3.61 & 11.81 & 34.19 & 5337242166699727232 & -0.12 $\pm$ 0.04 & -0.18 $\pm$ 0.04 & 0.21 $\pm$ 0.02 & 20.27 $\pm$ 7.05  \\
12820 & 11:12:04.15 & -60:44:48.36 & 4.46 & 4.70 & 12.93 & 37.95 & 5337636994459558016 & -0.01 $\pm$ 0.06 & 0.02 $\pm$ 0.06 & 0.02 $\pm$ 0.03 & 12.76 $\pm$ 34.91  \\
12821 & 11:10:32.91 & -61:12:06.90 & 4.61 & 4.12 & 11.46 & 33.14 & 5337244125204652160 & -0.15 $\pm$ 0.05 & 0.04 $\pm$ 0.04 & 0.15 $\pm$ 0.02 & 17.14 $\pm$ 9.38  \\
12827 & 11:09:46.16 & -61:26:19.00 & 4.60 & 4.87 & 8.93 & 39.89 & 5337218664636870400 & -0.51 $\pm$ 0.05 & -0.01 $\pm$ 0.05 & 0.51 $\pm$ 0.03 & 23.60 $\pm$ 3.15  \\
12829 & 11:11:00.36 & -61:04:48.68 & 4.47 & 4.29 & 11.95 & 31.61 & 5337250992822502912 & -0.18 $\pm$ 0.05 & 0.15 $\pm$ 0.05 & 0.23 $\pm$ 0.02 & 20.59 $\pm$ 6.75  \\
12841 & 11:10:41.23 & -61:11:11.13 & 4.51 & 3.77 & 11.98 & 32.27 & 5337244056485219072 & -0.22 $\pm$ 0.04 & 0.16 $\pm$ 0.04 & 0.27 $\pm$ 0.02 & 24.52 $\pm$ 4.37  \\
12847 & 11:10:56.47 & -61:07:09.35 & 4.58 & 4.56 & 11.24 & 31.32 & 5337250477426363264 & 0.23 $\pm$ 0.04 & -0.15 $\pm$ 0.04 & 0.28 $\pm$ 0.02 & 19.73 $\pm$ 4.97  \\
12862 & 11:10:44.82 & -61:11:46.10 & 4.52 & 3.92 & 11.11 & 31.76 & 5337244056485231360 & -0.21 $\pm$ 0.04 & -0.06 $\pm$ 0.04 & 0.22 $\pm$ 0.02 & 0.41 $\pm$ 7.12  \\
12884 & 11:10:34.23 & -61:16:21.47 & 4.58 & 4.16 & 11.97 & 32.76 & 5337241891821629184 & -0.02 $\pm$ 0.04 & -0.02 $\pm$ 0.04 & 0.03 $\pm$ 0.02 & 12.59 $\pm$ 28.55  \\
12911 & 11:11:18.22 & -61:06:29.10 & 4.51 & 4.76 & 12.08 & 29.02 & 5337250172519332864 & -0.10 $\pm$ 0.05 & 0.27 $\pm$ 0.05 & 0.29 $\pm$ 0.03 & 27.68 $\pm$ 1.99  \\
12913 & 11:11:05.34 & -61:10:32.63 & 4.58 & 4.05 & 11.34 & 29.50 & 5337243678528184704 & -0.62 $\pm$ 0.04 & -1.10 $\pm$ 0.04 & 1.26 $\pm$ 0.02 & 22.21 $\pm$ 0.84  \\
12917 & 11:12:17.17 & -60:49:21.69 & 4.51 & 6.59 & 10.57 & 33.34 & 5337624556233014912 & 0.79 $\pm$ 0.05 & 0.20 $\pm$ 0.05 & 0.81 $\pm$ 0.03 & 23.43 $\pm$ 1.79  \\
12921 & 11:10:45.46 & -61:17:18.93 & 4.57 & 3.78 & 12.88 & 31.44 & 5337241823102189056 & 0.03 $\pm$ 0.05 & 0.06 $\pm$ 0.05 & 0.07 $\pm$ 0.02 & 20.61 $\pm$ 18.65  \\
12935 & 11:11:35.25 & -61:04:00.09 & 4.47 & 5.90 & 11.51 & 28.03 & 5337251649988151168 & -0.23 $\pm$ 0.07 & 0.12 $\pm$ 0.08 & 0.26 $\pm$ 0.04 & 12.04 $\pm$ 9.82  \\
12945 & 11:12:11.29 & -60:54:02.54 & 4.52 & 8.25 & 11.82 & 30.27 & 5337617993522362496 & -0.28 $\pm$ 0.11 & 0.06 $\pm$ 0.11 & 0.29 $\pm$ 0.06 & 6.65 $\pm$ 13.00  \\
12953 & 11:11:50.93 & -61:00:45.33 & 4.59 & 7.14 & 11.11 & 27.93 & 5337240517431518208 & -0.27 $\pm$ 0.09 & 0.01 $\pm$ 0.09 & 0.27 $\pm$ 0.05 & 3.73 $\pm$ 9.39  \\
12967 & 11:09:49.61 & -61:37:56.28 & 4.58 & 3.90 & 11.72 & 43.97 & 5337206909279575296 & -0.22 $\pm$ 0.04 & -0.22 $\pm$ 0.04 & 0.31 $\pm$ 0.02 & 5.68 $\pm$ 8.32  \\
12977 & 11:12:03.85 & -60:59:12.85 & 4.56 & 7.13 & 12.03 & 27.52 & 5337615897578294528 & -0.20 $\pm$ 0.10 & 0.09 $\pm$ 0.09 & 0.22 $\pm$ 0.05 & 3.92 $\pm$ 14.68  \\
12990 & 11:13:48.80 & -60:27:56.58 & 4.46 & 7.16 & 12.48 & 48.63 & 5337658945999980672 & 2.10 $\pm$ 0.15 & -0.33 $\pm$ 0.12 & 2.12 $\pm$ 0.07 & 36.32 $\pm$ 2.18  \\
12992 & 11:10:56.23 & -61:20:18.53 & 4.62 & 6.76 & 10.69 & 30.44 & 5337240826669573504 & -0.90 $\pm$ 0.06 & -0.30 $\pm$ 0.06 & 0.95 $\pm$ 0.03 & 6.36 $\pm$ 2.36  \\
12994 & 11:10:08.69 & -61:34:20.94 & 4.47 & 3.97 & 12.30 & 40.26 & 5337208872110638208 & -0.27 $\pm$ 0.05 & -0.02 $\pm$ 0.05 & 0.27 $\pm$ 0.03 & 28.99 $\pm$ 5.00  \\
12997 & 11:12:39.90 & -60:49:44.34 & 4.50 & 5.90 & 10.15 & 31.41 & 5337624418794057344 & 0.04 $\pm$ 0.05 & 0.01 $\pm$ 0.05 & 0.05 $\pm$ 0.02 & 24.03 $\pm$ 19.03  \\
13005 & 11:12:34.24 & -60:52:13.14 & 4.50 & 8.73 & 11.54 & 29.81 & 5337618371479573760 & 1.44 $\pm$ 0.14 & 0.63 $\pm$ 0.13 & 1.58 $\pm$ 0.07 & 24.08 $\pm$ 1.93  \\
13015 & 11:11:01.32 & -61:20:35.69 & 4.57 & 7.59 & 10.57 & 29.88 & 5337240654870882048 & 0.02 $\pm$ 0.07 & -0.13 $\pm$ 0.08 & 0.13 $\pm$ 0.04 & 27.99 $\pm$ 6.49  \\
13037 & 11:10:22.28 & -61:33:54.80 & 4.62 & 4.93 & 11.22 & 38.63 & 5337208459793805696 & -0.79 $\pm$ 0.05 & 0.28 $\pm$ 0.04 & 0.84 $\pm$ 0.02 & 36.13 $\pm$ 0.90  \\
13047 & 11:10:31.08 & -61:32:34.96 & 4.64 & 4.79 & 11.35 & 37.09 & 5337208562873277056 & -0.41 $\pm$ 0.05 & 0.16 $\pm$ 0.04 & 0.44 $\pm$ 0.02 & 34.78 $\pm$ 1.64  \\
13051 & 11:11:25.74 & -61:16:48.39 & 4.56 & 7.73 & 12.73 & 26.59 & 5337230999783409024 & 0.20 $\pm$ 0.14 & 0.33 $\pm$ 0.13 & 0.38 $\pm$ 0.06 & 15.15 $\pm$ 10.17  \\
13057 & 11:11:42.33 & -61:12:30.11 & 4.46 & 5.39 & 12.48 & 24.79 & 5337231652618506496 & 0.24 $\pm$ 0.07 & -0.02 $\pm$ 0.06 & 0.24 $\pm$ 0.03 & 7.70 $\pm$ 6.21  \\
13061 & 11:12:37.07 & -60:56:12.96 & 4.57 & 7.66 & 11.04 & 26.54 & 5337617065771678592 & -0.21 $\pm$ 0.09 & 0.22 $\pm$ 0.09 & 0.30 $\pm$ 0.05 & 9.75 $\pm$ 10.07  \\
13076 & 11:13:03.07 & -60:49:25.53 & 4.52 & 5.20 & 12.14 & 30.18 & 5337623834678486656 & 0.01 $\pm$ 0.06 & 0.11 $\pm$ 0.06 & 0.11 $\pm$ 0.03 & 25.53 $\pm$ 8.29  \\
13088 & 11:13:04.05 & -60:50:25.00 & 4.54 & 7.53 & 12.20 & 29.27 & 5337623800281807744 & -0.20 $\pm$ 0.10 & 0.28 $\pm$ 0.11 & 0.34 $\pm$ 0.05 & 8.48 $\pm$ 11.71  \\
13089 & 11:12:15.74 & -61:05:04.81 & 4.49 & 7.12 & 10.81 & 23.16 & 5337240002035340288 & -0.34 $\pm$ 0.08 & 0.23 $\pm$ 0.07 & 0.41 $\pm$ 0.04 & 11.18 $\pm$ 5.12  \\
13093 & 11:12:38.05 & -60:58:47.65 & 4.54 & 7.03 & 10.72 & 24.62 & 5337615382182127232 & -0.09 $\pm$ 0.08 & 0.28 $\pm$ 0.08 & 0.29 $\pm$ 0.04 & 18.90 $\pm$ 5.49  \\
13097 & 11:13:04.68 & -60:51:02.87 & 4.58 & 5.77 & 11.83 & 28.68 & 5337623422361651584 & -0.09 $\pm$ 0.06 & 0.09 $\pm$ 0.06 & 0.13 $\pm$ 0.03 & 4.74 $\pm$ 17.80  \\
13098 & 11:10:40.50 & -61:34:02.57 & 4.55 & 4.41 & 12.09 & 36.79 & 5337208322354870656 & -0.24 $\pm$ 0.05 & -0.11 $\pm$ 0.05 & 0.26 $\pm$ 0.03 & 16.30 $\pm$ 8.15  \\
13099 & 11:11:53.97 & -61:12:40.93 & 4.51 & 5.90 & 11.71 & 23.38 & 5337237321975474304 & 0.05 $\pm$ 0.06 & 0.14 $\pm$ 0.06 & 0.15 $\pm$ 0.03 & 19.24 $\pm$ 6.12  \\
13118 & 11:13:40.38 & -60:43:13.40 & 4.54 & 4.68 & 10.81 & 34.06 & 5337646580827645056 & 0.17 $\pm$ 0.04 & 0.15 $\pm$ 0.04 & 0.22 $\pm$ 0.02 & 34.04 $\pm$ 0.35  \\
13126 & 11:12:42.08 & -61:01:50.69 & 4.44 & 5.46 & 12.31 & 22.25 & 5337614626267827200 & -0.15 $\pm$ 0.06 & 0.05 $\pm$ 0.06 & 0.16 $\pm$ 0.03 & 0.52 $\pm$ 9.82  \\
13129 & 11:11:02.93 & -61:31:26.80 & 4.51 & 4.00 & 11.95 & 33.19 & 5337202962235197696 & -0.36 $\pm$ 0.05 & 0.14 $\pm$ 0.04 & 0.39 $\pm$ 0.02 & 31.39 $\pm$ 1.57  \\
13158 & 11:10:38.94 & -61:41:08.76 & 4.45 & 4.11 & 12.40 & 40.90 & 5337205951532783488 & -0.30 $\pm$ 0.05 & 0.09 $\pm$ 0.05 & 0.31 $\pm$ 0.03 & 39.96 $\pm$ 1.67  \\
13161 & 11:12:25.39 & -61:09:52.60 & 4.50 & 4.82 & 10.32 & 20.26 & 5337237150176932352 & -0.08 $\pm$ 0.04 & -0.03 $\pm$ 0.04 & 0.08 $\pm$ 0.02 & 5.97 $\pm$ 11.10  \\
13178 & 11:12:23.19 & -61:12:18.09 & 4.44 & 5.58 & 12.17 & 19.96 & 5337236772219725312 & 0.16 $\pm$ 0.07 & -0.02 $\pm$ 0.06 & 0.16 $\pm$ 0.03 & 6.86 $\pm$ 7.80  \\
13187 & 11:12:26.02 & -61:12:00.83 & 4.53 & 5.46 & 12.48 & 19.68 & 5337236806579480192 & 0.14 $\pm$ 0.07 & 0.11 $\pm$ 0.06 & 0.18 $\pm$ 0.03 & 9.20 $\pm$ 8.29  \\
13198 & 11:12:27.83 & -61:12:24.53 & 4.57 & 5.32 & 12.09 & 19.39 & 5337236806579471616 & 0.08 $\pm$ 0.05 & -0.01 $\pm$ 0.05 & 0.08 $\pm$ 0.03 & 6.79 $\pm$ 11.75  \\
13204 & 11:13:14.37 & -60:58:59.15 & 4.53 & 7.49 & 12.51 & 21.48 & 5337613728578814976 & -0.22 $\pm$ 0.12 & 0.21 $\pm$ 0.10 & 0.30 $\pm$ 0.05 & 5.42 $\pm$ 10.12  \\
13209 & 11:11:10.66 & -61:37:00.61 & 4.49 & 4.39 & 12.51 & 35.40 & 5337196227726228480 & 0.19 $\pm$ 0.05 & -0.41 $\pm$ 0.05 & 0.45 $\pm$ 0.02 & 29.33 $\pm$ 2.88  \\
13212 & 11:11:22.13 & -61:33:43.61 & 4.62 & 3.83 & 10.92 & 32.39 & 5337202240680675200 & -1.96 $\pm$ 0.04 & -0.28 $\pm$ 0.04 & 1.98 $\pm$ 0.02 & 23.89 $\pm$ 0.49  \\
13218 & 11:11:43.33 & -61:28:54.00 & 4.56 & 6.60 & 11.02 & 27.74 & 5337203683789789824 & 0.05 $\pm$ 0.07 & 0.23 $\pm$ 0.06 & 0.24 $\pm$ 0.03 & 12.36 $\pm$ 8.49  \\
13220 & 11:14:36.25 & -60:36:25.96 & 4.44 & 4.12 & 12.00 & 39.38 & 5337653968171872640 & 0.16 $\pm$ 0.04 & 0.19 $\pm$ 0.04 & 0.25 $\pm$ 0.02 & 36.40 $\pm$ 3.65  \\
13225 & 11:12:54.71 & -61:08:33.81 & 4.44 & 4.40 & 11.65 & 17.40 & 5337235947586199296 & -0.10 $\pm$ 0.05 & -0.06 $\pm$ 0.04 & 0.12 $\pm$ 0.02 & 8.76 $\pm$ 7.11  \\
13244 & 11:12:53.12 & -61:12:11.67 & 4.58 & 5.12 & 10.02 & 16.44 & 5337235329110793472 & 0.19 $\pm$ 0.04 & -0.02 $\pm$ 0.04 & 0.20 $\pm$ 0.02 & 4.19 $\pm$ 3.56  \\
13273 & 11:13:21.01 & -61:08:27.84 & 4.46 & 4.92 & 11.52 & 14.62 & 5337610331300437248 & 0.13 $\pm$ 0.05 & -0.11 $\pm$ 0.04 & 0.17 $\pm$ 0.02 & 7.72 $\pm$ 4.49  \\
13280 & 11:13:42.89 & -61:03:07.60 & 4.58 & 6.81 & 11.50 & 16.09 & 5337612495964116096 & -1.14 $\pm$ 0.08 & 0.86 $\pm$ 0.08 & 1.43 $\pm$ 0.04 & 2.06 $\pm$ 1.20  \\
13285 & 11:13:12.38 & -61:13:05.94 & 4.54 & 4.41 & 12.23 & 13.99 & 5337235084263308544 & -0.33 $\pm$ 0.05 & 0.05 $\pm$ 0.04 & 0.34 $\pm$ 0.03 & 4.78 $\pm$ 2.02  \\
13288 & 11:15:34.82 & -60:29:09.83 & 4.56 & 4.13 & 11.21 & 46.60 & 5337561952809554944 & 0.91 $\pm$ 0.04 & 0.02 $\pm$ 0.04 & 0.91 $\pm$ 0.02 & 44.82 $\pm$ 0.53  \\
13305 & 11:14:25.65 & -60:54:17.61 & 4.49 & 6.61 & 10.92 & 21.91 & 5337433344286091648 & -0.51 $\pm$ 0.06 & -0.07 $\pm$ 0.06 & 0.52 $\pm$ 0.03 & 19.67 $\pm$ 1.36  \\
13325 & 11:13:34.84 & -61:13:22.12 & 4.52 & 4.46 & 11.57 & 11.29 & 5337234195239491200 & -0.17 $\pm$ 0.05 & -0.00 $\pm$ 0.04 & 0.17 $\pm$ 0.02 & 1.75 $\pm$ 2.52  \\
13329 & 11:13:51.97 & -61:08:29.86 & 4.51 & 5.70 & 10.30 & 11.49 & 5337610159501805952 & -0.36 $\pm$ 0.04 & 0.02 $\pm$ 0.04 & 0.36 $\pm$ 0.02 & 2.87 $\pm$ 1.22  \\
13331 & 11:13:51.48 & -61:08:49.08 & 4.50 & 6.49 & 10.33 & 11.34 & 5337610159501796736 & 0.04 $\pm$ 0.06 & 0.17 $\pm$ 0.05 & 0.18 $\pm$ 0.02 & 11.34 $\pm$ 0.08  \\
13333 & 11:13:32.81 & -61:15:01.18 & 4.46 & 5.32 & 11.26 & 11.31 & 5337234057800480128 & -0.11 $\pm$ 0.05 & 0.06 $\pm$ 0.04 & 0.13 $\pm$ 0.02 & 8.52 $\pm$ 3.33  \\
13341 & 11:13:36.34 & -61:15:03.60 & 4.54 & 4.64 & 12.57 & 10.88 & 5337233989081012480 & 0.18 $\pm$ 0.06 & -0.10 $\pm$ 0.05 & 0.21 $\pm$ 0.03 & 8.11 $\pm$ 2.55  \\
13346 & 11:13:38.28 & -61:14:43.65 & 4.48 & 4.30 & 12.47 & 10.68 & 5337234092160241152 & 0.04 $\pm$ 0.06 & 0.03 $\pm$ 0.05 & 0.05 $\pm$ 0.03 & 3.54 $\pm$ 10.07  \\
13347 & 11:15:43.60 & -60:36:01.87 & 4.45 & 4.10 & 12.70 & 39.86 & 5337467119913697024 & 0.07 $\pm$ 0.06 & 0.44 $\pm$ 0.06 & 0.45 $\pm$ 0.03 & 16.08 $\pm$ 5.47  \\
13362 & 11:13:46.30 & -61:16:01.11 & 4.63 & 5.01 & 10.75 & 9.68 & 5337046316178212864 & -0.77 $\pm$ 0.04 & -0.36 $\pm$ 0.04 & 0.85 $\pm$ 0.02 & 0.27 $\pm$ 0.62  \\
13364 & 11:13:52.24 & -61:14:23.10 & 4.61 & 4.32 & 11.60 & 9.05 & 5337046831574349568 & -0.29 $\pm$ 0.04 & -0.07 $\pm$ 0.04 & 0.30 $\pm$ 0.02 & 0.14 $\pm$ 1.46  \\
13366 & 11:13:20.21 & -61:24:28.77 & 4.60 & 8.57 & 11.26 & 15.54 & 5337225777103193728 & -0.18 $\pm$ 0.12 & -0.07 $\pm$ 0.09 & 0.19 $\pm$ 0.06 & 8.87 $\pm$ 8.02  \\
13369 & 11:12:26.89 & -61:41:03.50 & 4.46 & 7.14 & 11.72 & 31.79 & 5337197705195160320 & -0.41 $\pm$ 0.10 & 0.12 $\pm$ 0.08 & 0.43 $\pm$ 0.05 & 31.78 $\pm$ 0.16  \\
13377 & 11:15:15.37 & -60:51:17.75 & 4.57 & 5.26 & 9.12 & 24.36 & 5337456262235600512 & -0.36 $\pm$ 0.05 & 0.71 $\pm$ 0.05 & 0.80 $\pm$ 0.02 & 3.48 $\pm$ 1.97  \\
13390 & 11:14:28.78 & -61:07:46.21 & 4.51 & 9.84 & 9.71 & 9.10 & 5337424170235352448 & 0.51 $\pm$ 0.12 & 0.07 $\pm$ 0.10 & 0.52 $\pm$ 0.06 & 6.54 $\pm$ 1.38  \\
13392 & 11:14:15.74 & -61:12:11.57 & 4.56 & 6.17 & 11.23 & 7.04 & 5337422005571664640 & -0.10 $\pm$ 0.06 & 0.09 $\pm$ 0.05 & 0.13 $\pm$ 0.03 & 3.95 $\pm$ 3.36  \\
13396 & 11:14:05.80 & -61:15:41.93 & 4.63 & 5.78 & 11.85 & 7.33 & 5337046556696436864 & 0.13 $\pm$ 0.06 & -0.01 $\pm$ 0.05 & 0.13 $\pm$ 0.03 & 3.40 $\pm$ 2.74  \\
13399 & 11:12:57.29 & -61:36:31.87 & 4.58 & 7.64 & 10.56 & 26.00 & 5337198976505608192 & 0.01 $\pm$ 0.08 & -0.04 $\pm$ 0.07 & 0.04 $\pm$ 0.04 & 15.67 $\pm$ 9.22  \\
13407 & 11:14:10.24 & -61:15:45.62 & 4.61 & 5.48 & 11.93 & 6.79 & 5337046487976965632 & -0.21 $\pm$ 0.05 & -0.13 $\pm$ 0.05 & 0.24 $\pm$ 0.03 & 1.16 $\pm$ 1.81  \\
13415 & 11:14:10.08 & -61:17:26.31 & 4.63 & 6.40 & 11.61 & 7.04 & 5337046075660055808 & -0.07 $\pm$ 0.07 & -0.03 $\pm$ 0.06 & 0.08 $\pm$ 0.03 & 1.25 $\pm$ 6.17  \\
13416 & 11:14:46.42 & -61:06:18.82 & 4.53 & 5.82 & 11.98 & 9.64 & 5337423895357517184 & -0.07 $\pm$ 0.07 & 0.45 $\pm$ 0.06 & 0.46 $\pm$ 0.03 & 4.45 $\pm$ 1.37  \\
13422 & 11:14:26.63 & -61:13:25.19 & 4.49 & 5.67 & 11.99 & 5.31 & 5337421249657406208 & -0.21 $\pm$ 0.06 & -0.13 $\pm$ 0.05 & 0.25 $\pm$ 0.03 & 3.10 $\pm$ 1.38  \\
13426 & 11:15:00.02 & -61:03:45.51 & 4.48 & 5.15 & 12.30 & 11.91 & 5337429770873111936 & -0.27 $\pm$ 0.05 & -0.15 $\pm$ 0.05 & 0.31 $\pm$ 0.03 & 11.86 $\pm$ 0.23  \\
13427 & 11:14:19.04 & -61:16:26.90 & 4.48 & 5.99 & 12.75 & 5.79 & 5337045766422459008 & -0.18 $\pm$ 0.13 & 0.12 $\pm$ 0.09 & 0.22 $\pm$ 0.06 & 5.13 $\pm$ 1.67  \\
13436 & 11:13:57.71 & -61:24:26.82 & 4.64 & 6.93 & 9.85 & 12.09 & 5337043739197624064 & -0.34 $\pm$ 0.06 & -0.80 $\pm$ 0.05 & 0.87 $\pm$ 0.03 & 0.28 $\pm$ 1.05  \\
13437 & 11:14:32.30 & -61:13:55.90 & 4.62 & 7.68 & 10.69 & 4.48 & 5337421180937926400 & 0.05 $\pm$ 0.11 & -0.04 $\pm$ 0.08 & 0.06 $\pm$ 0.05 & 2.51 $\pm$ 3.35  \\
13440 & 11:14:40.58 & -61:11:28.27 & 4.52 & 6.65 & 12.34 & 5.22 & 5337421730693858176 & -0.10 $\pm$ 0.11 & -0.04 $\pm$ 0.08 & 0.11 $\pm$ 0.05 & 4.14 $\pm$ 2.78  \\
13443 & 11:14:47.71 & -61:09:40.55 & 4.47 & 6.49 & 12.46 & 6.39 & 5337423139443190272 & -0.35 $\pm$ 0.11 & 0.04 $\pm$ 0.08 & 0.36 $\pm$ 0.05 & 4.21 $\pm$ 1.21  \\
13448 & 11:14:20.26 & -61:18:24.75 & 4.59 & 7.46 & 12.16 & 6.24 & 5337045319745805824 & 0.20 $\pm$ 0.17 & -0.04 $\pm$ 0.11 & 0.21 $\pm$ 0.08 & 5.27 $\pm$ 2.11  \\
13452 & 11:16:46.39 & -60:33:27.48 & 4.46 & 4.39 & 12.02 & 43.88 & 5337557657842299008 & -0.15 $\pm$ 0.05 & 0.94 $\pm$ 0.04 & 0.95 $\pm$ 0.02 & 3.26 $\pm$ 2.38  \\
13459 & 11:15:11.39 & -61:04:08.58 & 4.64 & 5.46 & 11.97 & 11.51 & 5337426850295364480 & -0.21 $\pm$ 0.05 & -0.30 $\pm$ 0.05 & 0.37 $\pm$ 0.03 & 9.23 $\pm$ 1.33  \\
13461 & 11:14:59.76 & -61:08:34.72 & 4.51 & 6.62 & 11.14 & 7.11 & 5337423238186280960 & 0.30 $\pm$ 0.08 & 0.04 $\pm$ 0.07 & 0.31 $\pm$ 0.04 & 6.63 $\pm$ 0.64  \\
13465 & 11:14:25.01 & -61:20:02.64 & 4.64 & 8.44 & 10.07 & 6.67 & 5337045147947074432 & -0.20 $\pm$ 0.11 & 0.27 $\pm$ 0.09 & 0.34 $\pm$ 0.05 & 5.94 $\pm$ 1.28  \\
13466 & 11:14:25.61 & -61:20:05.50 & 4.59 & 7.65 & 11.36 & 6.65 & 5337045143634067072 & -0.32 $\pm$ 0.12 & 0.16 $\pm$ 0.09 & 0.36 $\pm$ 0.06 & 6.65 $\pm$ 0.03  \\
13467 & 11:14:17.34 & -61:22:36.94 & 4.57 & 6.09 & 10.29 & 9.16 & 5337043498679532160 & -0.04 $\pm$ 0.05 & 0.15 $\pm$ 0.04 & 0.16 $\pm$ 0.02 & 4.96 $\pm$ 2.89  \\
13471 & 11:14:45.51 & -61:15:00.20 & 4.62 & 7.16 & 9.04 & 2.63 & 5337420974779493248 & -0.12 $\pm$ 0.07 & -0.00 $\pm$ 0.06 & 0.12 $\pm$ 0.03 & 0.28 $\pm$ 1.19  \\
13472 & 11:14:46.08 & -61:14:50.64 & 4.52 & 6.99 & 12.66 & 2.61 & 5337421009139238912 & -0.25 $\pm$ 0.15 & -0.11 $\pm$ 0.11 & 0.27 $\pm$ 0.07 & 0.93 $\pm$ 1.35  \\
13486 & 11:15:00.35 & -61:12:29.87 & 4.64 & 6.63 & 10.58 & 3.23 & 5337418603957679104 & -0.13 $\pm$ 0.06 & 0.13 $\pm$ 0.05 & 0.19 $\pm$ 0.03 & 0.46 $\pm$ 1.32  \\
13490 & 11:14:57.32 & -61:14:39.25 & 4.62 & 7.56 & 11.06 & 1.50 & 5337418088561534848 & -0.11 $\pm$ 0.10 & -0.01 $\pm$ 0.07 & 0.11 $\pm$ 0.05 & 0.63 $\pm$ 0.90  \\
13492 & 11:14:55.25 & -61:15:20.61 & 4.63 & 5.29 & 10.37 & 1.41 & 5337418054201746432 & -0.08 $\pm$ 0.04 & 0.02 $\pm$ 0.04 & 0.08 $\pm$ 0.02 & 0.59 $\pm$ 0.72  \\
13493 & 11:14:52.88 & -61:16:07.11 & 4.62 & 5.39 & 12.34 & 1.74 & 5337417951098962560 & -0.11 $\pm$ 0.08 & 0.29 $\pm$ 0.07 & 0.31 $\pm$ 0.03 & 1.67 $\pm$ 0.15  \\
13494 & 11:14:56.27 & -61:15:12.23 & 4.62 & 5.97 & 11.26 & 1.33 & 5337418054201754752 & 0.19 $\pm$ 0.06 & 0.25 $\pm$ 0.05 & 0.31 $\pm$ 0.03 & 1.04 $\pm$ 0.20  \\
13495 & 11:13:29.60 & -61:41:35.96 & 4.51 & 5.84 & 12.96 & 28.44 & 5337010749528525440 & -0.63 $\pm$ 0.11 & -0.27 $\pm$ 0.08 & 0.68 $\pm$ 0.05 & 25.63 $\pm$ 2.15  \\
13500 & 11:15:48.45 & -60:59:15.38 & 4.59 & 5.12 & 12.13 & 17.14 & 5337429113701924736 & -0.08 $\pm$ 0.05 & 0.47 $\pm$ 0.05 & 0.47 $\pm$ 0.02 & 1.53 $\pm$ 2.07  \\
13503 & 11:15:02.30 & -61:13:51.94 & 4.60 & 7.54 & 10.94 & 1.85 & 5337418500878432128 & 0.07 $\pm$ 0.09 & 0.15 $\pm$ 0.07 & 0.17 $\pm$ 0.04 & 1.62 $\pm$ 0.56  \\
13509 & 11:15:00.69 & -61:15:01.80 & 4.62 & 6.16 & 10.65 & 0.95 & 5337418088538280064 & 0.07 $\pm$ 0.06 & 0.05 $\pm$ 0.04 & 0.09 $\pm$ 0.03 & 0.75 $\pm$ 0.41  \\
13511 & 11:14:59.78 & -61:15:24.67 & 4.64 & 5.25 & 11.35 & 0.87 & 5337418054202067584 & -0.05 $\pm$ 0.05 & -0.12 $\pm$ 0.05 & 0.12 $\pm$ 0.02 & 0.77 $\pm$ 0.20  \\
13512 & 11:15:01.35 & -61:15:01.21 & 4.64 & 6.56 & 11.34 & 0.89 & 5337418088561808000 & 0.10 $\pm$ 0.07 & 0.02 $\pm$ 0.06 & 0.10 $\pm$ 0.04 & 0.50 $\pm$ 0.49  \\
13516 & 11:15:00.31 & -61:15:30.35 & 4.47 & 5.20 & 12.25 & 0.78 & 5337418054201755776 & -0.23 $\pm$ 0.06 & 0.15 $\pm$ 0.05 & 0.27 $\pm$ 0.03 & 0.56 $\pm$ 0.16  \\
13517 & 11:14:59.87 & -61:15:43.60 & 4.52 & 6.29 & 12.32 & 0.83 & 5337418054201743744 & 0.18 $\pm$ 0.09 & -0.01 $\pm$ 0.07 & 0.18 $\pm$ 0.04 & 0.44 $\pm$ 0.29  \\
13519 & 11:15:02.38 & -61:15:07.73 & 4.54 & 6.11 & 12.73 & 0.73 & 5337418088561807872 & -0.74 $\pm$ 0.15 & 0.48 $\pm$ 0.11 & 0.89 $\pm$ 0.07 & 0.13 $\pm$ 0.14  \\
13520 & 11:13:52.29 & -61:36:31.28 & 4.63 & 6.06 & 11.63 & 22.70 & 5337034221549873792 & 0.09 $\pm$ 0.06 & -0.12 $\pm$ 0.06 & 0.15 $\pm$ 0.03 & 14.30 $\pm$ 9.12  \\
13521 & 11:15:04.09 & -61:14:39.60 & 4.60 & 6.85 & 12.34 & 1.03 & 5337418393470442752 & 0.11 $\pm$ 0.12 & 0.41 $\pm$ 0.09 & 0.43 $\pm$ 0.05 & 0.84 $\pm$ 0.19  \\
13522 & 11:15:10.92 & -61:12:50.18 & 4.47 & 7.24 & 11.91 & 2.84 & 5337418569582148608 & -0.18 $\pm$ 0.11 & 0.15 $\pm$ 0.08 & 0.23 $\pm$ 0.05 & 1.78 $\pm$ 1.18  \\
13526 & 11:15:04.97 & -61:14:58.87 & 4.65 & 7.92 & 10.91 & 0.69 & 5337418397799189248 & -0.13 $\pm$ 0.10 & -0.14 $\pm$ 0.07 & 0.19 $\pm$ 0.04 & 0.68 $\pm$ 0.05  \\
13529 & 11:15:03.97 & -61:15:35.63 & 4.64 & 4.78 & 11.22 & 0.34 & 5337418019842038912 & 0.00 $\pm$ 0.05 & -0.07 $\pm$ 0.04 & 0.07 $\pm$ 0.02 & 0.33 $\pm$ 0.04  \\
13530 & 11:15:07.71 & -61:14:35.47 & 4.63 & 7.22 & 11.72 & 1.05 & 5337418397776101376 & 0.22 $\pm$ 0.09 & 0.02 $\pm$ 0.07 & 0.22 $\pm$ 0.04 & 1.04 $\pm$ 0.06  \\
13532 & 11:15:13.36 & -61:13:03.50 & 4.55 & 7.60 & 11.43 & 2.70 & 5337418569580191872 & 0.12 $\pm$ 0.09 & 0.32 $\pm$ 0.08 & 0.34 $\pm$ 0.04 & 1.12 $\pm$ 0.79  \\
13533 & 11:15:05.66 & -61:15:26.93 & 4.58 & 5.04 & 11.51 & 0.23 & 5337418019842034432 & -0.02 $\pm$ 0.05 & 0.25 $\pm$ 0.04 & 0.25 $\pm$ 0.02 & 0.18 $\pm$ 0.03  \\
13534 & 11:15:11.51 & -61:13:38.67 & 4.48 & 7.79 & 12.32 & 2.07 & 5337418530903538432 & 0.04 $\pm$ 0.14 & 0.02 $\pm$ 0.11 & 0.04 $\pm$ 0.07 & 1.93 $\pm$ 0.56  \\
13541 & 11:15:06.61 & -61:15:23.71 & 4.59 & 4.92 & 11.27 & 0.24 & 5337418019842024192 & -0.00 $\pm$ 0.06 & -0.03 $\pm$ 0.05 & 0.03 $\pm$ 0.02 & 0.14 $\pm$ 0.18  \\
13542 & 11:13:25.90 & -61:45:57.88 & 4.60 & 6.19 & 12.80 & 32.63 & 5337008692248970496 & -0.71 $\pm$ 0.11 & -0.27 $\pm$ 0.08 & 0.76 $\pm$ 0.05 & 30.41 $\pm$ 1.75  \\
13546 & 11:15:07.82 & -61:15:27.84 & 4.60 & 5.19 & 9.95 & 0.21 & 5337418015513333888 & 0.02 $\pm$ 0.05 & 0.06 $\pm$ 0.04 & 0.06 $\pm$ 0.02 & 0.01 $\pm$ 0.18  \\
13549 & 11:15:06.66 & -61:15:52.68 & 4.59 & 5.42 & 11.91 & 0.24 & 5337418019842021120 & -0.22 $\pm$ 0.08 & -0.29 $\pm$ 0.05 & 0.36 $\pm$ 0.03 & 0.20 $\pm$ 0.04  \\
13550 & 11:15:13.52 & -61:13:46.95 & 4.60 & 8.21 & 10.59 & 2.02 & 5337418530902030720 & 0.23 $\pm$ 0.11 & 0.19 $\pm$ 0.09 & 0.29 $\pm$ 0.05 & 1.51 $\pm$ 0.61  \\
13555 & 11:15:08.90 & -61:15:27.22 & 4.62 & 4.78 & 10.93 & 0.32 & 5337418019842060928 & -0.07 $\pm$ 0.05 & 0.01 $\pm$ 0.04 & 0.07 $\pm$ 0.02 & 0.28 $\pm$ 0.09  \\
13556 & 11:15:11.55 & -61:14:40.48 & 4.61 & 7.65 & 11.22 & 1.12 & 5337418432158952576 & 0.02 $\pm$ 0.10 & -0.10 $\pm$ 0.07 & 0.10 $\pm$ 0.04 & 0.41 $\pm$ 0.90  \\
13560 & 11:15:08.19 & -61:15:47.21 & 4.61 & 5.12 & 9.64 & 0.23 & 5337418015513335552 & -0.02 $\pm$ 0.06 & -0.10 $\pm$ 0.05 & 0.10 $\pm$ 0.02 & 0.23 $\pm$ 0.02  \\
13567 & 11:15:18.03 & -61:12:54.28 & 4.58 & 7.16 & 9.28 & 3.05 & 5337419836572100096 & -0.08 $\pm$ 0.07 & 0.02 $\pm$ 0.06 & 0.08 $\pm$ 0.03 & 2.99 $\pm$ 0.46  \\
13570 & 11:15:09.84 & -61:15:30.42 & 4.51 & 5.07 & 9.08 & 0.39 & 5337418397799183872 & 0.26 $\pm$ 0.05 & 0.04 $\pm$ 0.05 & 0.27 $\pm$ 0.03 & 0.21 $\pm$ 0.07  \\
13571 & 11:15:08.08 & -61:16:03.37 & 4.63 & 5.29 & 10.86 & 0.45 & 5337418015513335168 & 0.12 $\pm$ 0.06 & -0.08 $\pm$ 0.04 & 0.14 $\pm$ 0.03 & 0.12 $\pm$ 0.19  \\
13572 & 11:15:06.69 & -61:16:33.32 & 4.62 & 5.69 & 9.86 & 0.92 & 5337417985482256768 & -0.14 $\pm$ 0.04 & -0.11 $\pm$ 0.04 & 0.18 $\pm$ 0.02 & 0.87 $\pm$ 0.09  \\
13573 & 11:15:17.11 & -61:13:21.50 & 4.53 & 6.89 & 12.20 & 2.60 & 5337418363439522432 & 0.38 $\pm$ 0.13 & -0.51 $\pm$ 0.09 & 0.64 $\pm$ 0.05 & 1.81 $\pm$ 0.45  \\
13574 & 11:15:10.86 & -61:15:17.68 & 4.46 & 6.01 & 12.51 & 0.60 & 5337418393470468864 & -0.03 $\pm$ 0.11 & -0.05 $\pm$ 0.07 & 0.06 $\pm$ 0.04 & 0.02 $\pm$ 0.54  \\
13576 & 11:15:08.71 & -61:15:59.85 & 4.55 & 5.60 & 8.68 & 0.43 & 5337418015513337472 & -0.34 $\pm$ 0.06 & -0.03 $\pm$ 0.05 & 0.34 $\pm$ 0.03 & 0.28 $\pm$ 0.05  \\
13578 & 11:15:19.09 & -61:12:48.90 & 4.58 & 6.23 & 11.53 & 3.19 & 5337419840908289664 & 0.15 $\pm$ 0.06 & 0.23 $\pm$ 0.05 & 0.27 $\pm$ 0.03 & 1.39 $\pm$ 0.81  \\
13579 & 11:15:10.07 & -61:15:37.99 & 4.61 & 5.11 & 9.64 & 0.40 & 5337418019842057088 & -0.15 $\pm$ 0.06 & -0.35 $\pm$ 0.05 & 0.38 $\pm$ 0.03 & 0.28 $\pm$ 0.05  \\
13580 & 11:15:11.52 & -61:15:16.85 & 4.57 & 6.19 & 10.18 & 0.67 & 5337418191640765184 & 0.14 $\pm$ 0.06 & 0.04 $\pm$ 0.04 & 0.14 $\pm$ 0.03 & 0.40 $\pm$ 0.21  \\
13581 & 11:15:09.35 & -61:16:01.97 & 4.63 & 5.09 & 9.19 & 0.51 & 5337418015513340160 & -0.16 $\pm$ 0.06 & -0.02 $\pm$ 0.05 & 0.16 $\pm$ 0.03 & 0.30 $\pm$ 0.14  \\
13582 & 11:15:07.10 & -61:16:44.39 & 4.65 & 7.41 & 11.31 & 1.10 & 5337417985482251904 & 0.08 $\pm$ 0.10 & -0.46 $\pm$ 0.08 & 0.47 $\pm$ 0.04 & 0.27 $\pm$ 0.25  \\
13584 & 11:15:11.07 & -61:15:36.81 & 4.61 & 4.96 & 11.26 & 0.52 & 5337418191640752896 & 0.00 $\pm$ 0.05 & -0.07 $\pm$ 0.04 & 0.07 $\pm$ 0.02 & 0.48 $\pm$ 0.12  \\
13585 & 11:15:21.37 & -61:12:28.34 & 4.61 & 6.76 & 10.37 & 3.62 & 5337419870931846272 & 0.02 $\pm$ 0.05 & 0.23 $\pm$ 0.05 & 0.23 $\pm$ 0.02 & 0.19 $\pm$ 0.90  \\
13587 & 11:15:12.24 & -61:15:25.50 & 4.57 & 5.60 & 11.48 & 0.69 & 5337418191640762112 & -0.10 $\pm$ 0.05 & 0.13 $\pm$ 0.04 & 0.16 $\pm$ 0.02 & 0.69 $\pm$ 0.00  \\
13589 & 11:15:07.58 & -61:16:54.66 & 4.64 & 7.16 & 9.24 & 1.28 & 5337417985482248576 & 0.06 $\pm$ 0.05 & 0.08 $\pm$ 0.05 & 0.10 $\pm$ 0.02 & 1.12 $\pm$ 0.42  \\
13590 & 11:15:11.39 & -61:15:45.45 & 4.61 & 4.64 & 11.57 & 0.57 & 5337417813683625856 & 0.01 $\pm$ 0.05 & 0.43 $\pm$ 0.05 & 0.43 $\pm$ 0.03 & 0.56 $\pm$ 0.01  \\
13592 & 11:15:14.51 & -61:14:53.88 & 4.63 & 6.32 & 11.30 & 1.19 & 5337418226000522240 & 0.02 $\pm$ 0.07 & 0.15 $\pm$ 0.05 & 0.15 $\pm$ 0.03 & 0.47 $\pm$ 0.52  \\
13594 & 11:15:11.32 & -61:15:55.60 & 4.63 & 4.54 & 10.51 & 0.62 & 5337417813683617152 & 0.10 $\pm$ 0.04 & 0.23 $\pm$ 0.04 & 0.25 $\pm$ 0.02 & 0.60 $\pm$ 0.04  \\
13596 & 11:15:16.24 & -61:14:25.36 & 4.48 & 7.74 & 12.77 & 1.67 & 5337418329079753088 & 0.18 $\pm$ 0.21 & -0.01 $\pm$ 0.14 & 0.18 $\pm$ 0.10 & 1.58 $\pm$ 0.39  \\
13597 & 11:15:15.83 & -61:14:38.22 & 4.60 & 7.53 & 12.52 & 1.48 & 5337418226000531712 & 0.18 $\pm$ 0.14 & -0.19 $\pm$ 0.11 & 0.26 $\pm$ 0.06 & 1.40 $\pm$ 0.30  \\
13601 & 11:15:10.66 & -61:16:19.94 & 4.57 & 5.10 & 10.78 & 0.84 & 5337417809354915584 & 0.22 $\pm$ 0.04 & 0.02 $\pm$ 0.04 & 0.22 $\pm$ 0.02 & 0.55 $\pm$ 0.12  \\
13603 & 11:15:12.76 & -61:15:44.81 & 4.57 & 5.00 & 12.28 & 0.73 & 5337418191640752384 & 0.32 $\pm$ 0.06 & 0.09 $\pm$ 0.06 & 0.33 $\pm$ 0.03 & 0.05 $\pm$ 0.15  \\
13609 & 11:14:11.56 & -61:34:48.03 & 4.48 & 6.94 & 12.96 & 20.27 & 5337034496427836160 & -0.51 $\pm$ 0.14 & -0.40 $\pm$ 0.13 & 0.65 $\pm$ 0.07 & 16.48 $\pm$ 3.42  \\
13611 & 11:15:12.90 & -61:16:06.79 & 4.61 & 5.31 & 11.13 & 0.88 & 5337417813683615872 & -0.07 $\pm$ 0.05 & 0.08 $\pm$ 0.04 & 0.11 $\pm$ 0.02 & 0.51 $\pm$ 0.40  \\
13613 & 11:15:09.57 & -61:17:14.01 & 4.46 & 7.05 & 12.00 & 1.63 & 5337417676244596736 & -0.06 $\pm$ 0.10 & -0.13 $\pm$ 0.08 & 0.14 $\pm$ 0.04 & 1.40 $\pm$ 0.66  \\
13615 & 11:15:17.61 & -61:14:52.80 & 4.59 & 6.34 & 11.55 & 1.51 & 5337418226000528512 & -0.15 $\pm$ 0.07 & 0.04 $\pm$ 0.06 & 0.16 $\pm$ 0.04 & 1.36 $\pm$ 0.31  \\
13617 & 11:15:13.83 & -61:16:03.76 & 4.48 & 5.13 & 12.82 & 0.95 & 5337417813660359680 & 0.17 $\pm$ 0.08 & -0.14 $\pm$ 0.08 & 0.23 $\pm$ 0.04 & 0.54 $\pm$ 0.36  \\
13619 & 11:15:18.09 & -61:14:56.93 & 4.56 & 6.57 & 12.06 & 1.53 & 5337418226000528000 & -0.18 $\pm$ 0.09 & -0.00 $\pm$ 0.07 & 0.18 $\pm$ 0.04 & 1.12 $\pm$ 0.38  \\
13623 & 11:15:13.83 & -61:16:29.33 & 4.59 & 5.92 & 11.26 & 1.20 & 5337417813660246784 & -0.13 $\pm$ 0.05 & -0.02 $\pm$ 0.04 & 0.13 $\pm$ 0.02 & 0.67 $\pm$ 0.38  \\
13625 & 11:15:16.18 & -61:15:57.77 & 4.62 & 5.58 & 11.29 & 1.18 & 5337418191640754432 & 0.02 $\pm$ 0.05 & -0.09 $\pm$ 0.04 & 0.09 $\pm$ 0.02 & 1.17 $\pm$ 0.08  \\
13629 & 11:15:01.65 & -61:20:41.88 & 4.55 & 9.90 & 9.57 & 5.10 & 5337042364808329216 & 0.42 $\pm$ 0.13 & 0.22 $\pm$ 0.10 & 0.48 $\pm$ 0.06 & 4.96 $\pm$ 0.37  \\
13631 & 11:15:15.68 & -61:16:27.37 & 4.47 & 5.82 & 12.74 & 1.35 & 5337417809354932224 & 0.13 $\pm$ 0.10 & 0.08 $\pm$ 0.07 & 0.15 $\pm$ 0.05 & 1.02 $\pm$ 0.61  \\
13633 & 11:15:18.91 & -61:15:28.63 & 4.54 & 5.68 & 11.68 & 1.47 & 5337418226000514944 & 0.11 $\pm$ 0.06 & 0.17 $\pm$ 0.05 & 0.20 $\pm$ 0.03 & 0.76 $\pm$ 0.46  \\
13637 & 11:15:24.17 & -61:14:00.25 & 4.59 & 7.31 & 11.53 & 2.66 & 5337418294720042240 & 0.25 $\pm$ 0.10 & 0.09 $\pm$ 0.07 & 0.26 $\pm$ 0.05 & 1.68 $\pm$ 0.80  \\
13638 & 11:15:23.75 & -61:14:10.31 & 4.61 & 7.37 & 12.65 & 2.51 & 5337418294720036096 & 0.17 $\pm$ 0.14 & 0.04 $\pm$ 0.10 & 0.18 $\pm$ 0.07 & 1.75 $\pm$ 1.18  \\
13640 & 11:15:22.50 & -61:14:35.59 & 4.58 & 7.49 & 12.04 & 2.16 & 5337418260360283776 & -0.14 $\pm$ 0.11 & -0.05 $\pm$ 0.08 & 0.15 $\pm$ 0.06 & 1.15 $\pm$ 1.26  \\
13642 & 11:15:21.33 & -61:14:58.09 & 4.57 & 7.02 & 12.55 & 1.88 & 5337418260360272384 & -0.18 $\pm$ 0.12 & -0.01 $\pm$ 0.09 & 0.18 $\pm$ 0.06 & 1.22 $\pm$ 0.72  \\
13643 & 11:15:21.33 & -61:15:04.36 & 4.62 & 6.76 & 9.60 & 1.84 & 5337418157281054336 & 0.00 $\pm$ 0.06 & 0.11 $\pm$ 0.05 & 0.11 $\pm$ 0.02 & 1.42 $\pm$ 0.66  \\
13648 & 11:15:21.91 & -61:15:01.33 & 4.59 & 6.85 & 11.83 & 1.92 & 5337418260360272000 & 0.21 $\pm$ 0.10 & 0.05 $\pm$ 0.08 & 0.22 $\pm$ 0.05 & 0.85 $\pm$ 0.75  \\
13649 & 11:15:22.45 & -61:14:58.55 & 4.59 & 6.40 & 11.54 & 2.00 & 5337418260337262976 & 0.07 $\pm$ 0.07 & 0.13 $\pm$ 0.07 & 0.15 $\pm$ 0.04 & 0.74 $\pm$ 1.11  \\
13657 & 11:15:29.10 & -61:13:14.79 & 4.46 & 5.57 & 12.80 & 3.60 & 5337419806525824128 & 0.23 $\pm$ 0.09 & 0.24 $\pm$ 0.08 & 0.33 $\pm$ 0.04 & 1.03 $\pm$ 1.24  \\
13659 & 11:15:24.08 & -61:14:49.83 & 4.45 & 5.86 & 12.30 & 2.23 & 5337418260360280576 & 0.16 $\pm$ 0.09 & 0.11 $\pm$ 0.07 & 0.19 $\pm$ 0.04 & 0.26 $\pm$ 1.21  \\
13662 & 11:15:19.38 & -61:16:19.33 & 4.54 & 5.62 & 12.59 & 1.67 & 5337418122921272704 & -0.10 $\pm$ 0.09 & -0.02 $\pm$ 0.07 & 0.10 $\pm$ 0.04 & 0.44 $\pm$ 1.25  \\
13663 & 11:15:35.53 & -61:11:23.52 & 4.56 & 6.38 & 11.51 & 5.48 & 5337419566030445056 & 0.09 $\pm$ 0.06 & 0.05 $\pm$ 0.06 & 0.10 $\pm$ 0.03 & 3.73 $\pm$ 2.90  \\
13664 & 11:14:53.55 & -61:24:22.89 & 4.62 & 6.40 & 9.73 & 8.89 & 5337040337583630208 & -0.25 $\pm$ 0.06 & -0.08 $\pm$ 0.05 & 0.26 $\pm$ 0.03 & 8.83 $\pm$ 0.23  \\
13666 & 11:15:29.61 & -61:13:36.33 & 4.49 & 5.84 & 12.46 & 3.42 & 5337419767860057088 & 0.06 $\pm$ 0.07 & 0.16 $\pm$ 0.06 & 0.18 $\pm$ 0.03 & 0.69 $\pm$ 1.56  \\
13667 & 11:15:22.69 & -61:15:52.08 & 4.58 & 5.52 & 12.25 & 1.93 & 5337418152952337280 & 0.08 $\pm$ 0.07 & 0.03 $\pm$ 0.05 & 0.08 $\pm$ 0.03 & 0.22 $\pm$ 1.49  \\
13674 & 11:15:33.29 & -61:12:57.43 & 4.51 & 5.71 & 11.56 & 4.17 & 5337419046305558912 & -0.04 $\pm$ 0.05 & 0.09 $\pm$ 0.04 & 0.10 $\pm$ 0.02 & 3.18 $\pm$ 1.58  \\
13676 & 11:15:24.17 & -61:15:57.52 & 4.60 & 6.54 & 11.29 & 2.12 & 5337418157257919360 & 0.12 $\pm$ 0.07 & -0.05 $\pm$ 0.07 & 0.13 $\pm$ 0.03 & 1.17 $\pm$ 1.09  \\
13677 & 11:15:33.46 & -61:13:05.25 & 4.47 & 5.77 & 12.57 & 4.10 & 5337419050634324864 & -0.01 $\pm$ 0.07 & 0.20 $\pm$ 0.06 & 0.20 $\pm$ 0.03 & 2.28 $\pm$ 1.22  \\
13679 & 11:15:21.80 & -61:16:53.58 & 4.59 & 5.70 & 10.98 & 2.20 & 5337417744964134400 & 0.04 $\pm$ 0.05 & 0.14 $\pm$ 0.06 & 0.14 $\pm$ 0.03 & 2.20 $\pm$ 0.03  \\
13683 & 11:15:29.88 & -61:14:26.98 & 4.54 & 5.73 & 12.56 & 3.02 & 5337418294720041472 & -0.20 $\pm$ 0.07 & -0.01 $\pm$ 0.06 & 0.20 $\pm$ 0.04 & 1.97 $\pm$ 0.74  \\
13690 & 11:15:18.03 & -61:18:40.59 & 4.59 & 7.69 & 9.32 & 3.33 & 5337417607525100544 & -0.09 $\pm$ 0.08 & 0.23 $\pm$ 0.08 & 0.24 $\pm$ 0.04 & 1.31 $\pm$ 1.22  \\
13692 & 11:15:55.62 & -61:07:04.19 & 4.54 & 4.30 & 10.52 & 10.39 & 5337425716424009344 & -0.31 $\pm$ 0.04 & -0.16 $\pm$ 0.04 & 0.35 $\pm$ 0.02 & 7.89 $\pm$ 1.00  \\
13694 & 11:15:27.94 & -61:15:50.66 & 4.61 & 5.62 & 10.06 & 2.56 & 5337417397038104832 & -0.27 $\pm$ 0.04 & -0.02 $\pm$ 0.04 & 0.27 $\pm$ 0.02 & 0.55 $\pm$ 0.41  \\
13695 & 11:15:28.82 & -61:15:36.13 & 4.56 & 6.09 & 12.07 & 2.65 & 5337417504446026368 & 0.11 $\pm$ 0.07 & -0.16 $\pm$ 0.06 & 0.19 $\pm$ 0.03 & 2.58 $\pm$ 0.29  \\
13696 & 11:15:02.97 & -61:23:36.54 & 4.60 & 6.59 & 11.88 & 7.99 & 5337041643253735424 & -0.30 $\pm$ 0.08 & -0.13 $\pm$ 0.07 & 0.33 $\pm$ 0.04 & 7.95 $\pm$ 0.22  \\
13697 & 11:15:28.73 & -61:15:41.22 & 4.60 & 5.96 & 12.33 & 2.64 & 5337417397038106624 & 0.12 $\pm$ 0.09 & 0.18 $\pm$ 0.09 & 0.22 $\pm$ 0.04 & 1.52 $\pm$ 1.17  \\
13703 & 11:14:01.50 & -61:42:27.05 & 4.57 & 4.91 & 11.21 & 27.92 & 5337009860494168832 & -0.72 $\pm$ 0.04 & -0.28 $\pm$ 0.04 & 0.77 $\pm$ 0.02 & 26.78 $\pm$ 0.54  \\
13704 & 11:15:29.74 & -61:15:34.73 & 4.56 & 6.48 & 12.90 & 2.76 & 5337417500117324928 & -0.25 $\pm$ 0.12 & -0.17 $\pm$ 0.10 & 0.31 $\pm$ 0.06 & 0.56 $\pm$ 1.33  \\
13708 & 11:15:25.78 & -61:17:04.30 & 4.50 & 5.88 & 12.56 & 2.70 & 5337417740635483904 & -0.59 $\pm$ 0.13 & -1.14 $\pm$ 0.13 & 1.28 $\pm$ 0.06 & 2.59 $\pm$ 0.10  \\
13709 & 11:15:34.19 & -61:14:29.00 & 4.44 & 5.85 & 12.53 & 3.50 & 5337419016274555264 & 0.18 $\pm$ 0.08 & 0.06 $\pm$ 0.07 & 0.19 $\pm$ 0.04 & 1.28 $\pm$ 1.59  \\
13717 & 11:15:32.80 & -61:15:36.99 & 4.55 & 7.28 & 11.82 & 3.13 & 5337417504446033408 & 0.32 $\pm$ 0.11 & 0.05 $\pm$ 0.09 & 0.33 $\pm$ 0.05 & 0.65 $\pm$ 1.02  \\
13718 & 11:15:27.73 & -61:17:13.77 & 4.64 & 6.31 & 12.30 & 2.98 & 5337416989049894912 & -0.17 $\pm$ 0.09 & 0.09 $\pm$ 0.07 & 0.19 $\pm$ 0.04 & 0.82 $\pm$ 1.50  \\
13722 & 11:15:36.16 & -61:15:04.09 & 4.58 & 6.66 & 11.40 & 3.58 & 5337417538805793024 & 0.20 $\pm$ 0.08 & -0.02 $\pm$ 0.07 & 0.21 $\pm$ 0.04 & 2.17 $\pm$ 1.09  \\
13725 & 11:15:06.54 & -61:24:41.16 & 4.60 & 6.10 & 11.79 & 9.05 & 5337041540174491008 & -0.19 $\pm$ 0.07 & 0.01 $\pm$ 0.06 & 0.20 $\pm$ 0.04 & 8.19 $\pm$ 1.26  \\
13726 & 11:15:06.33 & -61:24:50.43 & 4.58 & 7.31 & 12.92 & 9.20 & 5337041535855293568 & -0.24 $\pm$ 0.17 & 0.28 $\pm$ 0.13 & 0.37 $\pm$ 0.07 & 3.10 $\pm$ 4.80  \\
13728 & 11:15:43.81 & -61:13:36.62 & 4.50 & 7.04 & 10.81 & 4.90 & 5337419084994069760 & 0.12 $\pm$ 0.08 & 0.22 $\pm$ 0.08 & 0.25 $\pm$ 0.04 & 1.41 $\pm$ 2.07  \\
13731 & 11:15:43.06 & -61:14:12.35 & 4.55 & 8.44 & 11.44 & 4.59 & 5337418981915015680 & 0.33 $\pm$ 0.12 & 0.11 $\pm$ 0.10 & 0.35 $\pm$ 0.06 & 1.58 $\pm$ 1.66  \\
13738 & 11:14:09.45 & -61:43:53.04 & 4.59 & 6.13 & 12.20 & 29.06 & 5337006832519070208 & -0.06 $\pm$ 0.07 & 0.04 $\pm$ 0.07 & 0.07 $\pm$ 0.03 & 22.67 $\pm$ 17.20  \\
13739 & 11:15:34.83 & -61:17:56.23 & 4.63 & 7.73 & 11.52 & 4.08 & 5337416920330395264 & 0.09 $\pm$ 0.11 & 0.09 $\pm$ 0.09 & 0.13 $\pm$ 0.05 & 3.44 $\pm$ 1.94  \\
13740 & 11:15:40.34 & -61:16:18.05 & 4.50 & 6.89 & 12.44 & 4.09 & 5337417435726552832 & 0.17 $\pm$ 0.12 & 0.25 $\pm$ 0.09 & 0.31 $\pm$ 0.05 & 2.81 $\pm$ 1.45  \\
13744 & 11:15:08.07 & -61:26:54.58 & 4.56 & 6.57 & 11.75 & 11.27 & 5337039925266728576 & 0.28 $\pm$ 0.09 & -0.29 $\pm$ 0.07 & 0.41 $\pm$ 0.04 & 4.15 $\pm$ 2.81  \\
13754 & 11:15:50.49 & -61:15:53.83 & 4.45 & 6.98 & 12.14 & 5.26 & 5337418741396643456 & 0.26 $\pm$ 0.10 & -0.11 $\pm$ 0.09 & 0.29 $\pm$ 0.05 & 3.42 $\pm$ 1.68  \\
13761 & 11:15:45.81 & -61:17:57.36 & 4.58 & 7.57 & 11.63 & 5.23 & 5337417126489134336 & 0.37 $\pm$ 0.11 & 0.33 $\pm$ 0.10 & 0.50 $\pm$ 0.05 & 3.80 $\pm$ 1.04  \\
13765 & 11:15:40.54 & -61:19:49.81 & 4.55 & 8.41 & 7.62 & 5.84 & 5337416679812175232 & 0.14 $\pm$ 0.09 & 0.26 $\pm$ 0.08 & 0.30 $\pm$ 0.04 & 5.83 $\pm$ 0.11  \\
13766 & 11:15:43.32 & -61:18:59.08 & 4.44 & 6.12 & 12.07 & 5.52 & 5337416714171947264 & -0.86 $\pm$ 0.07 & -0.28 $\pm$ 0.07 & 0.91 $\pm$ 0.04 & 3.12 $\pm$ 0.46  \\
13770 & 11:15:33.45 & -61:22:33.15 & 4.57 & 6.04 & 12.52 & 7.62 & 5337041402735673856 & -0.04 $\pm$ 0.09 & 0.02 $\pm$ 0.07 & 0.05 $\pm$ 0.04 & 1.46 $\pm$ 5.13  \\
13790 & 11:16:04.42 & -61:15:11.37 & 4.55 & 6.24 & 11.88 & 6.95 & 5337407059084440960 & 0.10 $\pm$ 0.07 & 0.13 $\pm$ 0.07 & 0.16 $\pm$ 0.03 & 3.24 $\pm$ 3.35  \\
13793 & 11:14:33.70 & -61:43:28.79 & 4.59 & 6.88 & 11.86 & 28.12 & 5337007008635909120 & 0.13 $\pm$ 0.09 & -0.14 $\pm$ 0.08 & 0.20 $\pm$ 0.04 & 13.57 $\pm$ 14.21  \\
13794 & 11:14:21.95 & -61:47:01.98 & 4.50 & 8.68 & 10.58 & 31.85 & 5337005153209975296 & -0.05 $\pm$ 0.12 & -0.11 $\pm$ 0.11 & 0.13 $\pm$ 0.06 & 18.78 $\pm$ 24.29  \\
13799 & 11:16:02.82 & -61:16:52.01 & 4.49 & 6.27 & 12.29 & 6.85 & 5337406917314140544 & 0.20 $\pm$ 0.08 & 0.15 $\pm$ 0.07 & 0.25 $\pm$ 0.04 & 2.92 $\pm$ 2.54  \\
13804 & 11:15:16.83 & -61:32:37.37 & 4.49 & 7.07 & 12.37 & 17.03 & 5337032641002080640 & 0.49 $\pm$ 0.14 & -0.96 $\pm$ 0.11 & 1.07 $\pm$ 0.06 & 0.56 $\pm$ 2.76  \\
13834 & 11:16:56.34 & -61:05:56.25 & 4.58 & 4.97 & 12.35 & 16.39 & 5337415167983277952 & -0.20 $\pm$ 0.06 & -0.01 $\pm$ 0.06 & 0.20 $\pm$ 0.03 & 13.26 $\pm$ 2.83  \\
13839 & 11:17:49.25 & -60:50:41.68 & 4.57 & 7.21 & 12.55 & 31.76 & 5337444889161636096 & -0.27 $\pm$ 0.11 & 1.65 $\pm$ 0.10 & 1.67 $\pm$ 0.05 & 14.23 $\pm$ 2.15  \\
13860 & 11:16:50.16 & -61:13:14.22 & 4.59 & 8.05 & 10.11 & 12.67 & 5337412487923505792 & 0.76 $\pm$ 0.08 & 0.47 $\pm$ 0.08 & 0.89 $\pm$ 0.04 & 0.03 $\pm$ 1.61  \\
13885 & 11:17:46.65 & -61:01:03.42 & 4.51 & 5.16 & 11.19 & 24.18 & 5337437639256186880 & -0.15 $\pm$ 0.05 & -0.13 $\pm$ 0.04 & 0.19 $\pm$ 0.02 & 7.11 $\pm$ 7.53  \\
13888 & 11:16:49.27 & -61:19:28.96 & 4.44 & 6.62 & 12.40 & 12.90 & 5337405890853499648 & 0.37 $\pm$ 0.10 & 0.10 $\pm$ 0.10 & 0.38 $\pm$ 0.05 & 2.64 $\pm$ 3.90  \\
13892 & 11:17:07.78 & -61:14:30.84 & 4.56 & 4.86 & 12.01 & 14.60 & 5337412213045599488 & 0.07 $\pm$ 0.05 & -0.04 $\pm$ 0.05 & 0.08 $\pm$ 0.02 & 12.03 $\pm$ 5.81  \\
13893 & 11:16:10.21 & -61:32:33.71 & 4.53 & 6.92 & 12.95 & 18.55 & 5337026726807903616 & 0.43 $\pm$ 0.14 & -0.24 $\pm$ 0.12 & 0.49 $\pm$ 0.07 & 4.96 $\pm$ 5.99  \\
13902 & 11:16:40.98 & -61:24:20.95 & 4.55 & 7.61 & 9.35 & 14.27 & 5337403726189732224 & 0.10 $\pm$ 0.08 & 0.17 $\pm$ 0.07 & 0.20 $\pm$ 0.04 & 13.82 $\pm$ 1.72  \\
13908 & 11:16:38.10 & -61:26:42.85 & 4.49 & 6.21 & 11.81 & 15.58 & 5337403554391015424 & 1.13 $\pm$ 0.07 & -0.58 $\pm$ 0.06 & 1.27 $\pm$ 0.03 & 0.79 $\pm$ 1.06  \\
13910 & 11:15:49.29 & -61:42:02.37 & 4.47 & 6.99 & 12.22 & 26.89 & 5337018072473741312 & 0.11 $\pm$ 0.11 & 0.08 $\pm$ 0.10 & 0.13 $\pm$ 0.05 & 26.78 $\pm$ 2.11  \\
13912 & 11:15:49.24 & -61:42:06.38 & 4.55 & 7.10 & 11.13 & 26.95 & 5337018072473447808 & 0.12 $\pm$ 0.09 & -0.18 $\pm$ 0.09 & 0.21 $\pm$ 0.04 & 0.42 $\pm$ 14.56  \\
13918 & 11:16:33.08 & -61:29:49.40 & 4.60 & 6.78 & 10.34 & 17.55 & 5337027693199975040 & 1.37 $\pm$ 0.07 & -0.85 $\pm$ 0.06 & 1.62 $\pm$ 0.03 & 0.30 $\pm$ 0.93  \\
13931 & 11:17:12.93 & -61:20:08.59 & 4.59 & 6.57 & 10.92 & 15.81 & 5337403073354793088 & 1.58 $\pm$ 0.07 & 0.10 $\pm$ 0.06 & 1.58 $\pm$ 0.03 & 0.25 $\pm$ 0.64  \\
13934 & 11:16:02.90 & -61:42:20.66 & 4.50 & 7.50 & 12.54 & 27.54 & 5337018794027963136 & 0.17 $\pm$ 0.15 & 0.16 $\pm$ 0.12 & 0.23 $\pm$ 0.07 & 27.16 $\pm$ 3.37  \\
13937 & 11:15:56.65 & -61:44:43.52 & 4.55 & 7.71 & 11.81 & 29.69 & 5337017179120227328 & 0.00 $\pm$ 0.15 & -0.38 $\pm$ 0.11 & 0.38 $\pm$ 0.06 & 15.92 $\pm$ 9.29  \\
13939 & 11:16:03.30 & -61:43:12.64 & 4.50 & 6.38 & 12.20 & 28.39 & 5337017316559202304 & 0.28 $\pm$ 0.08 & -0.05 $\pm$ 0.08 & 0.29 $\pm$ 0.04 & 20.27 $\pm$ 6.19  \\
13942 & 11:16:02.19 & -61:43:50.53 & 4.52 & 6.70 & 10.98 & 28.97 & 5337017282199455744 & 0.28 $\pm$ 0.08 & -0.19 $\pm$ 0.07 & 0.33 $\pm$ 0.04 & 10.36 $\pm$ 8.55  \\
13945 & 11:15:59.30 & -61:45:21.04 & 4.53 & 8.09 & 11.23 & 30.37 & 5337017144760806016 & 0.16 $\pm$ 0.12 & -0.36 $\pm$ 0.11 & 0.39 $\pm$ 0.06 & 4.96 $\pm$ 11.41  \\
13946 & 11:15:49.45 & -61:48:42.23 & 4.56 & 7.56 & 11.94 & 33.46 & 5336993127301868928 & -0.22 $\pm$ 0.14 & -0.01 $\pm$ 0.12 & 0.23 $\pm$ 0.07 & 29.96 $\pm$ 8.17  \\
13953 & 11:17:01.32 & -61:27:55.66 & 4.62 & 6.97 & 10.70 & 18.43 & 5337400358935538432 & 0.35 $\pm$ 0.07 & -0.04 $\pm$ 0.06 & 0.35 $\pm$ 0.04 & 4.27 $\pm$ 3.67  \\
13954 & 11:16:12.62 & -61:43:54.25 & 4.62 & 7.84 & 9.48 & 29.34 & 5337017247839725184 & 0.10 $\pm$ 0.09 & -0.13 $\pm$ 0.09 & 0.16 $\pm$ 0.05 & 0.16 $\pm$ 20.79  \\
13966 & 11:16:22.54 & -61:43:42.17 & 4.44 & 5.17 & 10.79 & 29.49 & 5337018484790318464 & -0.55 $\pm$ 0.04 & -0.43 $\pm$ 0.04 & 0.69 $\pm$ 0.02 & 29.48 $\pm$ 0.05  \\
13974 & 11:17:18.38 & -61:27:38.30 & 4.60 & 6.74 & 12.69 & 19.82 & 5337400221496603392 & 0.10 $\pm$ 0.12 & 0.12 $\pm$ 0.10 & 0.16 $\pm$ 0.05 & 18.08 $\pm$ 6.77  \\
13978 & 11:19:09.18 & -60:52:48.75 & 4.47 & 6.74 & 12.98 & 37.15 & 5337398606592603008 & -0.17 $\pm$ 0.13 & -0.10 $\pm$ 0.12 & 0.19 $\pm$ 0.06 & 17.54 $\pm$ 24.70  \\
13997 & 11:17:31.35 & -61:27:49.95 & 4.47 & 6.23 & 12.42 & 21.19 & 5337399774801917312 & 0.05 $\pm$ 0.09 & -0.12 $\pm$ 0.10 & 0.12 $\pm$ 0.05 & 17.12 $\pm$ 10.06  \\
14015 & 11:17:31.89 & -61:30:03.34 & 4.48 & 6.14 & 12.33 & 22.58 & 5337399362503126400 & 0.32 $\pm$ 0.08 & 0.20 $\pm$ 0.08 & 0.38 $\pm$ 0.04 & 17.38 $\pm$ 4.03  \\
14016 & 11:17:07.64 & -61:37:41.26 & 4.61 & 8.09 & 10.61 & 26.36 & 5337022333081110016 & 0.16 $\pm$ 0.10 & 0.19 $\pm$ 0.10 & 0.25 $\pm$ 0.05 & 26.27 $\pm$ 1.20  \\
14034 & 11:18:14.23 & -61:20:15.38 & 4.46 & 4.26 & 12.76 & 22.98 & 5337407849359101568 & 0.04 $\pm$ 0.05 & -0.09 $\pm$ 0.05 & 0.10 $\pm$ 0.03 & 22.14 $\pm$ 3.91  \\
14052 & 11:17:39.68 & -61:33:12.68 & 4.54 & 6.00 & 12.71 & 25.37 & 5337023634431445888 & 0.11 $\pm$ 0.10 & 0.05 $\pm$ 0.08 & 0.12 $\pm$ 0.05 & 18.52 $\pm$ 14.41  \\
14053 & 11:19:31.77 & -60:57:14.01 & 4.50 & 5.64 & 11.23 & 36.92 & 5337396991685074560 & -0.19 $\pm$ 0.06 & 0.08 $\pm$ 0.06 & 0.21 $\pm$ 0.03 & 35.54 $\pm$ 3.43  \\
14079 & 11:18:07.51 & -61:28:35.73 & 4.49 & 5.18 & 11.64 & 25.24 & 5337354145085586560 & 0.30 $\pm$ 0.05 & -0.05 $\pm$ 0.05 & 0.30 $\pm$ 0.03 & 0.41 $\pm$ 4.75  \\
14080 & 11:18:18.28 & -61:25:24.14 & 4.48 & 4.50 & 11.88 & 24.95 & 5337354656151746048 & 0.14 $\pm$ 0.05 & 0.01 $\pm$ 0.04 & 0.14 $\pm$ 0.02 & 2.06 $\pm$ 8.24  \\
14092 & 11:17:48.79 & -61:36:14.58 & 4.52 & 7.05 & 10.14 & 28.28 & 5337023398233319552 & 0.05 $\pm$ 0.07 & -0.30 $\pm$ 0.07 & 0.31 $\pm$ 0.03 & 22.97 $\pm$ 4.33  \\
14112 & 11:20:03.20 & -60:55:12.25 & 4.57 & 4.52 & 9.68 & 41.24 & 5337394444726197632 & 0.28 $\pm$ 0.06 & -0.06 $\pm$ 0.06 & 0.29 $\pm$ 0.03 & 36.28 $\pm$ 4.56  \\
14156 & 11:18:48.38 & -61:24:53.53 & 4.55 & 5.68 & 12.00 & 28.14 & 5337359677004293120 & 0.24 $\pm$ 0.06 & 0.07 $\pm$ 0.06 & 0.25 $\pm$ 0.03 & 7.25 $\pm$ 7.56  \\
14157 & 11:18:43.10 & -61:26:35.87 & 4.44 & 6.89 & 10.82 & 28.15 & 5337353732769778176 & 0.20 $\pm$ 0.06 & -0.05 $\pm$ 0.06 & 0.21 $\pm$ 0.03 & 5.83 $\pm$ 9.75  \\
14167 & 11:18:40.58 & -61:28:33.30 & 4.56 & 6.31 & 9.95 & 28.69 & 5337353595330856704 & 0.08 $\pm$ 0.05 & -0.10 $\pm$ 0.05 & 0.13 $\pm$ 0.02 & 20.96 $\pm$ 9.96  \\
14172 & 11:19:21.21 & -61:16:03.86 & 4.56 & 4.31 & 12.28 & 30.59 & 5337385751715379840 & 0.59 $\pm$ 0.04 & 0.22 $\pm$ 0.05 & 0.63 $\pm$ 0.02 & 0.03 $\pm$ 2.86  \\
14182 & 11:19:53.64 & -61:06:24.13 & 4.61 & 6.70 & 12.62 & 35.78 & 5337391459767058304 & 0.39 $\pm$ 0.13 & 0.43 $\pm$ 0.12 & 0.58 $\pm$ 0.06 & 7.34 $\pm$ 10.49  \\
14183 & 11:18:50.52 & -61:27:04.86 & 4.50 & 5.19 & 11.28 & 29.16 & 5337353664050307584 & -0.71 $\pm$ 0.04 & -0.56 $\pm$ 0.05 & 0.90 $\pm$ 0.02 & 18.88 $\pm$ 1.64  \\
14203 & 11:19:00.19 & -61:26:31.71 & 4.45 & 6.47 & 10.68 & 30.03 & 5337359367766864000 & 0.14 $\pm$ 0.06 & -0.15 $\pm$ 0.06 & 0.21 $\pm$ 0.03 & 21.81 $\pm$ 8.20  \\
14215 & 11:20:29.69 & -60:59:09.58 & 4.60 & 7.14 & 12.45 & 42.33 & 5337393212113907456 & -0.18 $\pm$ 0.12 & -0.07 $\pm$ 0.11 & 0.19 $\pm$ 0.06 & 16.42 $\pm$ 26.04  \\
14238 & 11:20:34.17 & -60:59:37.19 & 4.53 & 6.06 & 11.37 & 42.65 & 5337393212113904512 & 0.12 $\pm$ 0.08 & 0.17 $\pm$ 0.07 & 0.21 $\pm$ 0.04 & 8.12 $\pm$ 19.66  \\
14248 & 11:19:08.98 & -61:28:33.57 & 4.47 & 6.38 & 11.79 & 31.77 & 5337353423532261120 & 0.19 $\pm$ 0.07 & -0.26 $\pm$ 0.09 & 0.32 $\pm$ 0.04 & 24.88 $\pm$ 6.60  \\
14257 & 11:20:25.06 & -61:04:37.64 & 4.45 & 5.03 & 12.79 & 39.92 & 5337391116169983232 & -0.00 $\pm$ 0.07 & -0.35 $\pm$ 0.07 & 0.35 $\pm$ 0.03 & 31.94 $\pm$ 4.81  \\
14281 & 11:19:55.12 & -61:16:03.69 & 4.55 & 5.66 & 9.66 & 34.66 & 5337383007271110656 & 0.32 $\pm$ 0.04 & -0.25 $\pm$ 0.04 & 0.41 $\pm$ 0.02 & 29.49 $\pm$ 2.71  \\
14304 & 11:19:35.16 & -61:28:28.11 & 4.46 & 8.82 & 10.97 & 34.62 & 5337356206670950400 & 0.22 $\pm$ 0.12 & -0.12 $\pm$ 0.11 & 0.25 $\pm$ 0.06 & 15.83 $\pm$ 17.46  \\
14359 & 11:20:45.20 & -61:16:27.84 & 4.59 & 4.06 & 10.27 & 40.68 & 5337371771635501824 & -0.75 $\pm$ 0.05 & -0.34 $\pm$ 0.05 & 0.82 $\pm$ 0.03 & 3.69 $\pm$ 3.28  \\
14365 & 11:22:11.34 & -60:48:57.86 & 4.51 & 2.79 & 11.20 & 57.90 & 5337483195934871680 & -0.45 $\pm$ 0.05 & -0.73 $\pm$ 0.05 & 0.85 $\pm$ 0.02 & 10.38 $\pm$ 4.57  \\
14370 & 11:21:52.32 & -60:55:44.07 & 4.46 & 5.57 & 11.90 & 52.89 & 5337476293974024832 & 0.08 $\pm$ 0.07 & 0.07 $\pm$ 0.06 & 0.10 $\pm$ 0.03 & 0.66 $\pm$ 42.72  \\
14391 & 11:21:24.80 & -61:08:31.90 & 4.49 & 5.19 & 10.27 & 46.08 & 5337379433857665024 & 0.25 $\pm$ 0.04 & -0.24 $\pm$ 0.04 & 0.35 $\pm$ 0.02 & 44.23 $\pm$ 2.07  \\
 \hline
\label{tab:crossref}\\
 \end{longtable}
 \end{landscape}

}
\twocolumn

\label{lastpage}

\end{document}